\documentclass[aps,pra,twocolumn,superscriptaddress,%raggedbottom,%showpacs,
floatfix]{revtex4-1}
\usepackage{amsmath}
\usepackage[next]{inputenc}
\usepackage[dvips]{epsfig}
\usepackage{bbm,bm,bbold}
\usepackage{booktabs}
\usepackage{multirow}
\usepackage{hhline}
\usepackage{comment}
\usepackage{float}

\usepackage{amsmath,amsfonts,amssymb,amsthm}
\usepackage{color}
\usepackage{graphicx,latexsym}

\usepackage[colorlinks=true,urlcolor=blue,citecolor=blue,linkcolor=blue]{hyperref}

\usepackage{lipsum}

\usepackage{nameref}
\usepackage{varioref}
\usepackage{cleveref}

\usepackage{natbib}

\begin{document}
	\renewcommand{\vec}{\mathbf}
	\renewcommand{\Re}{\mathop{\mathrm{Re}}\nolimits}
	\renewcommand{\Im}{\mathop{\mathrm{Im}}\nolimits}

\preprint{APS/123-QED}

\title{Electron-exciton interactions in the exciton-polaron problem}% Force line breaks with \\

\author{Dmitry K. Efimkin}
 \email{dmitry.efimkin@monash.edu}
 \affiliation{School of Physics and Astronomy and ARC Centre of Excellence in Future Low-Energy Electronics Technologies, Monash University, Victoria 3800, Australia }

 \author{Emma K. Laird}%
 
 \affiliation{School of Physics and Astronomy and ARC Centre of Excellence in Future Low-Energy Electronics Technologies, Monash University, Victoria 3800, Australia }

 \author{Jesper Levinsen}%
 
 \affiliation{School of Physics and Astronomy and ARC Centre of Excellence in Future Low-Energy Electronics Technologies, Monash University, Victoria 3800, Australia }
 
\author{Meera M.~Parish}%
\affiliation{School of Physics and Astronomy and ARC Centre of Excellence in Future Low-Energy Electronics Technologies, Monash University, Victoria 3800, Australia }

	\author{Allan H. MacDonald}
\affiliation{The Center for Complex Quantum Systems, The University of Texas at Austin, Austin, Texas 78712-1192, USA}


\begin{abstract}
Recently, it has been demonstrated that the absorption of moderately doped two-dimensional semiconductors can be described in terms of exciton-polarons. In this scenario, attractive and repulsive polaron branches are formed due to interactions between a photo-excited exciton and a Fermi sea of excess charge carriers. These interactions have previously been treated in a phenomenological manner. Here, we present a microscopic derivation of the electron-exciton interactions which utilizes a mixture of variational and perturbative approaches. We find that the interactions feature classical charge-dipole behavior in the long-range limit, and that they are only weakly modified for moderate doping. We apply our theory to the absorption properties and show that the dependence on doping is well captured by a model with a phenomenological contact potential.  
%splitting between polaron branches is well captured by the phenomenological contact potential; however the redistribution of weights (oscillator strengths) between branches is considerably underestimated. Thus, we conclude that the finite range of interactions is important.

%The absorption of moderately doped two-dimensional semiconductors can be described in terms of exciton-polarons. Attractive and repulsive exciton-polarons are formed due to interactions between a photo-excited exciton and Fermi sea of excess charge carriers. Previously these interactions have been treated in a phenomenological way, while the present paper presents their microscopic derivation with the help of variational and perturbative approaches. The resulting interactions have the classical charge-dipole behavior at large separations between exciton and electron. We have compered doping dependence of the absorption with our previous results and have found that splitting between peaks and their broadening are well captured by the phenomenological contact potential. However, we find that redistribution of weights (oscillator strengths) between the polaronic branches are considerably underestimated allowing us to conclude that the finite range of interactions is important. 
\end{abstract}

\maketitle

%\tableofcontents

\section{Introduction}
\label{SecI}
%A decade ago  
The fabrication of graphene has opened the door to the world of \emph{flatland materials}~\cite{Graphene1,Graphene2, GrapheneMacDonald}.
This diverse family is still growing and prominently includes the transition-metal dichalcogenide (TMDC) monolayers, $\mathrm{MoS}_2$, $\mathrm{MoSe}_2$, $\mathrm{WS}_2$ and $\mathrm{WSe}_2$. These are two-dimensional semiconductors featuring a direct bandgap and extraordinarily strong Coulomb interactions, such that their optical properties are dominated by exciton physics (bound electron-hole pairs) even at room temperature. An important %peculiarity 
property of TMDC monolayers is a strong spin-valley splitting. Together with the possibility of valley selective light-matter coupling by using circularly polarized light, this opens avenues for spintronics and valleytronics~\cite{ReviewValleytronics1,ReviewValleytronics2}. The ability to combine TMDC monolayers in lateral and vertical heterostructures furthermore makes them %very promising 
strong candidates for optoelectronic applications (\textit{e.g.},~\cite{Applications1,Applications2,Applications3,Applications4,Applications5}), and %creates 
promises
a %vast
versatile platform for the exploration of exciton physics~\cite{ExcitonReview1,ExcitonReview2,ExcitonReviewGlazov}. 

An additional advantage of TMDC monolayers is that their optical properties can be tuned by gating.  The presence of excess charge carriers (electrons or holes) has been found~\cite{TrionExperiment1,TrionExperiment2,TrionExperiment3,TrionExperiment4,TrionExperiment5,TrionEnergyRecent,TrionEnergyRecent} to split the exciton feature in absorption spectra~\footnote{One should note that photoluminescence %(PL) 
experiments with ultraclean TMDC samples have also reported the presence of additional peaks, which have been attributed to biexcitons, localized excitons and trions, as well as to four- and five-particle electron-hole complexes (\textit{e.g.},~\cite{ManyParticleComplexes1,ManyParticleComplexes2})} into two peaks~\footnote{It should be noted that this phenomenon has been observed in conventional semiconductor GaAs and CdTe quantum wells (QWs)~\cite{QW1,QW2,QW3,QW4,QW5}. However, the binding energies, $\epsilon_\mathrm{X}$ and $\epsilon_\mathrm{T}$, in QWs are more than an order of magnitude smaller than in TMDC monolayers, and this restricts the observability of phenomena in QWs to very low temperatures.}. The presence of the %additional 
redshifted peak has conventionally %usually 
been attributed to trions, charged and weakly-coupled three-particle complexes formed by binding two electrons to one hole, or two holes to one electron~\cite{ExcitonReview1,ExcitonReview2,ExcitonReviewGlazov}. Its binding energy $\epsilon_\mathrm{T}$ is interpreted to be equal to the splitting between peaks in the limit of vanishing doping~\cite{TrionTheory1,TrionTheory2,TrionTheory3, TrionTheory4,XPVagov1,XPVagov2}. 
However, recent work~\cite{TMDCDemlerExp,EfimkinMacDonald1,EfimkinMacDonald2} 
%(see also Ref.~\cite{TMDCDemlerExp}) we have 
has argued that the three-particle picture for the additional peak is only relevant at very low doping and cannot explain the doping dependence of absorption (see also earlier works~\cite{Suris1,Suris2,Wouters1,Wouters2, Combescot1, Combescot2, Combescot3,Zimmermann} where the trion scenario has been questioned).
Instead, within the wide doping range where the excess-carrier Fermi energy $\epsilon_\mathrm{F}\sim \epsilon_\mathrm{T}$, %with $\epsilon_\mathrm{F}$ to be the Fermi energy of excess charge carriers 
the appropriate picture is one of excitons interacting with the degenerate Fermi sea of excess charge carriers. In this case, the excitons are dressed by excitations of the Fermi sea, forming attractive and repulsive exciton-polaron (XP) quasiparticles 
like in the Fermi-polaron problem introduced in the context of cold atoms~\cite{FermiPolaronReview1,Levinsen2015review}.
%~\cite{FermiPolaron1, FermiPolaron2,FermiPolaron4, FermiPolaron3, FermiPolaronNew3, FermiPolaronNew2,FermiPolaronDemler,  FermiPolaron5,FermiPolaronNew1}, 
%the interactions split excitons into attractive and repulsive exciton-polaron (XP) quasiparticles. 
These quasiparticles represent a many-body generalization of exciton-electron (X-e) bound and unbound states. 

The theory of absorption by XPs developed  in Refs.~\cite{EfimkinMacDonald1,EfimkinMacDonald2} has naturally and successfully explained its observed doping dependence (it should be mentioned that there is an alternative picture that is based on dynamical screening and exciton-plasmon interactions~\cite{Dery1,Dery2,Dery3}). However, there are still a number of experimentally relevant open questions that are within the focus of recent theoretical research~\cite{XPnew1,XPnew2,GalzovXT,ImamogluSchmidt,XPVagov1,XPVagov2}. These include the manifestations of the XP physics in photoluminescence (PL) and in nonlinear optical phenomena~\cite{XPnonlinear,TrionNonlinear1,TrionNonlinear2,bastarracheamagnani2020attractive}, as well as the crossover between few-particle and Fermi polaron physics. For the latter problem, it has been recently argued~\cite{GalzovXT} that predictions of XP and trion based absorption theories are almost indistinguishable in the limit of low doping. Another outstanding question concerns the importance of the microscopic details of the X-e interactions, which have been approximated by a phenomenological contact potential in previous work~\cite{EfimkinMacDonald1,EfimkinMacDonald2,TMDCDemlerExp}. These details are also essential if excitons and excess charge carriers are spatially separated into different layers, a scenario that opens new avenues to control the flow of excitons~\cite{PolaronDragExp,PolaronDragTheory}. 
 
In the present work, we have used a mixture of %the help
variational and perturbative techniques to derive the microscopic exciton-electron interactions. We have found that these interactions possess a weak doping dependence.  Furthermore, they can be well approximated by a local potential which depends only on the X-e distance, and which features classical charge-dipole behavior at large separation. We find that this potential compares well to those recently obtained by other researchers~\cite{XPnew1,ImamogluSchmidt}, and that the resulting energy of the exciton-electron bound state agrees well with previous numerical results for the trion binding energy in the full three-body problem~\cite{TrionEnergy1,TrionEnergy2,TrionEnergyRecent}. We compare our calculated doping dependence of the optical conductivity with that obtained in a model with a phenomenological contact potential, and find that these agree very well. This indicates that including the finite range of exciton-electron interactions is not essential to capturing XP physics. 

%(responsible for absorption) is presented in Fig.~\ref{FigOpticalConductivity}, where it is compared to previous results obtained for the phenomenological contact potential~\cite{EfimkinMacDonald1}. We see that the doping dependence of both the splitting between attractive and repulsive XPs, and the line broadenings, are well approximated by the contact interactions. However, the redistribution of the spectral (oscillator) weights with doping is strongly underestimated by the theory with a phenomenological contact potential.  This indicates that including the finite range of interactions is essential to capturing the XP physics. 

This paper is organized as follows: In  Sec.~\ref{SecII}, we introduce our model for describing the optical properties of TMDC monolayers. We briefly review the phenomenological theory for XPs, and then we present our novel microscopic theory in Sec.~\ref{sec:SecXPTheory}. We discuss the details of our microscopically derived X-e interactions in Sec.~\ref{Sec:X-e}, and we also compare them to the recent results of other investigations. In Sec.~\ref{SecXPApsorption}, we present a microscopic calculation of the optical conductivity and we describe its dependence on frequency and doping.  Section~\ref{SecConc} contains discussions and a brief summary.  

\section{Model}
\label{SecII}

\begin{figure}[t]
	\begin{center}
		\includegraphics[trim=2cm 13cm 7.5cm 1cm, clip, width=1.0\columnwidth]{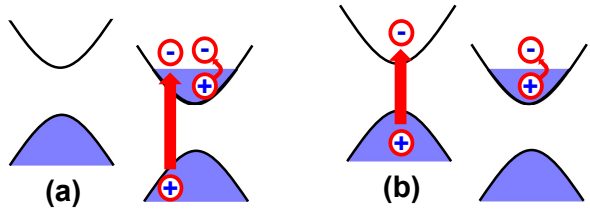}
		\caption{Low-energy region of the TMDC band structure in the presence of strong valley imbalance. Selective coupling with circularly polarized light leads to the creation of exciton-polarons dressed solely by either the indistinguishable (a) or the  distinguishable (b) Fermi sea. The second band arrangement (b) is considered in the present work.
    	\label{FigBands}
		}
	\end{center}
\end{figure}

The low-energy electron and hole states in TMDC monolayers are concentrated near two equivalent valleys ($K$ and $K'$) and are separated by %the direct gap $\epsilon_\mathrm{g}$. 
a direct bandgap. In undoped monolayers, the absorption spectrum is dominated by excitons that can be selectively excited in one of the valleys by circularly polarized light. In the presence of excess charge carriers (we assume these to be electrons with the generalization to holes straightforward), the valleys are generically equally populated which provides two Fermi seas (FSs) for the polaronic dressing of excitons. Importantly, electrons in one valley are distinguishable from the photo-excited electron hidden in the exciton, while electrons in the other valley are clearly not. The complicated interplay between the exchange and the polaronic physics with two FSs still represents a challenge for the microscopic theory of absorption. However, the effects of two FSs can be disentangled in the presence of strong valley splitting which depletes electrons in one of the two valleys. Since the $K$ and $K'$ points are time reversal partners, the splitting can be induced by exchange coupling to an insulating magnetic substrate~\cite{MagneticSubstarte} or by a magnetic field~\cite{MagneticField}. 

Two possible scenarios in the presence of strong splitting are sketched in Fig.~\ref{FigBands}. %(a) and~\ref{FigBands}(b).
Figure~\ref{FigBands}(a) illustrates the case where the photo-excited electron is indistinguishable from the electrons in the Fermi sea. In this case, the polaronic dressing is limited by exchange physics and momentum-space restrictions (%that are 
usually referred to as the Pauli-blocking effect)~\cite{Parish2011,Cotlet2020}. It is firmly established experimentally in the TMDCs~\cite{MagneticField,TMDCMF1,TMDCMF2,TMDCMF3,TMDCMF4} that this does not result in a splitting of the excitonic feature in absorption~\footnote{The polaronic effect leads to a splitting only when there is an electron-hole mass imbalance~\cite{TrionEnergyRecent}.  However, this imbalance is removed for TMDC monolayers and also in the presence of a magnetic field~\cite{QWMF1,QWMF2,QWMF3}}. While at the theoretical level the role of exchange physics on the absorption peaks is understood only at very low doping~\cite{TrionEnergyRecent}, this problem is outside the scope of the present work. The dressing by a distinguishable Fermi sea is sketched in Fig.~\ref{FigBands}(b). Due to the absence of both exchange physics and the Pauli-blocking effect, this regime is more favorable for polaronic physics, and it is the scenario considered in the current work.

The physics of XPs for the band arrangement shown in Fig.~\ref{FigBands}(b) can be described by the Hamiltonian, $H=H_0+H_\mathrm{C}+H_\mathrm{LM}$, which includes the kinetic energy $H_0$, Coulomb interactions $H_\mathrm{C}$, and light-matter interactions $H_\mathrm{LM}$. The %former
first of these is given by 
\begin{equation}
\label{H0}
H_0=\sum_\vec{p}\Big[\epsilon_\vec{p}^\mathrm{e} e_\vec{p}^\dagger e_\vec{p} + \epsilon_\vec{p}^\mathrm{h} h_\vec{p}^\dagger h_\vec{p} + \epsilon_\vec{p}^\mathrm{f}  f_\vec{p}^\dagger f_\vec{p}\Big]. 
\end{equation}
Here, $e_\vec{p}^\dagger$ and $h_\vec{p}^\dagger$ are the creation operators for photo-excited electrons and holes, respectively, with $f_\vec{p}^\dagger$ the operator for excess electrons. The band masses for electrons and holes in monolayer TMDCs can be well approximated as equal to each other.  Thus, we take the dispersions to be $\epsilon_\vec{p}^{\mathrm{e}(\mathrm{h})}=\epsilon_\mathrm{g}/2+p^2/2m$ (with $p\equiv|\vec{p}|$) and $\epsilon_\vec{p}^\mathrm{f}=p^2/2m-\epsilon_\mathrm{F}$, where $\epsilon_\mathrm{g}$ is the gap in the spectrum and $\epsilon_\mathrm{F}$ is the Fermi energy of excess charge carriers.  Note that, in this work, we employ Gaussian units ($4\pi\varepsilon_0=1$) and we additionally set the system area to unity.

%Neglecting intra-species interactions~\footnote{The exciton-polaron state consists of a single photo-excited electron-hole pair, which is why electron-electron $H_\mathrm{ee}$ and hole-hole  $H_\mathrm{hh}$ interactions are irrelevant for absorption and might be important only for the nonlinear optical effects that are outside the scope of the present work.}, the remaining interactions can be described by the following Hamiltonian 
The Coulombic interactions can be modelled by
%
%\begin{equation}
\begin{align}
\label{eq:HC}
H_\mathrm{C}=
\sum_{\vec{p} \bar{\vec{p}}\vec{q}} U_\vec{q} \Big[ e_{\vec{p}+\vec{q}}^\dagger f_{\bar{\vec{p}}-\vec{q}}^\dagger f_{\bar{\vec{p}}} e_\vec{p} &- h_{\vec{p}+\vec{q}}^\dagger f_{\bar{\vec{p}}-\vec{q}}^\dagger f_{\bar{\vec{p}}} h_\vec{p}   \nonumber\\&- e_{\vec{p}+\vec{q}}^\dagger h_{\bar{\vec{p}}-\vec{q}}^\dagger h_{\bar{\vec{p}}} e_\vec{p} \Big],   
\end{align}
%\end{equation}
%
where we neglect intraspecies interactions since these do not significantly affect the absorption~\footnote{The exciton-polaron state comprises only a single photo-excited electron-hole pair.  As a result, both electron-electron and hole-hole interactions are irrelevant for absorption and might only be important for nonlinear optical effects that are outside the scope of this work.}.
Here, $U_\vec{q}$ is the Coulomb interaction potential and our developed microscopic theory for exciton-polarons does not rely on its explicit form. For numerical calculations, we employ the two-dimensional (2D) Coulomb potential, $U_\vec{q}=2\pi e^2/\kappa q$, with $\kappa$ the effective dielectric constant of the surrounding media. However, we argue that our results are much more general and can be extended to other screening models, including the Keldysh potential~\cite{Keldysh1,Keldysh2,Keldysh3}. The Keldysh potential properly handles dielectric screening and is argued to quite accurately capture the spectrum of excitonic states in TMDC monolayers~\cite{TMDCEx1,TrionExperiment4}. 

Within the dipole approximation, light-matter interactions can be described by using a position-independent vector potential, $\vec{A}$, as follows: 
\begin{equation}
\label{HLM}
H_\mathrm{LM}= - \frac{e v }{c} \sum_\vec{p} \vec{A}\cdot \Big[\vec{e} \; e^\dagger_\vec{p} h^\dagger_{-\vec{p}}e^{- i \omega t} + \mathrm{h.c.} \Big].  \\  
\end{equation}
The light frequency satisfies $\omega\sim\epsilon_\mathrm{g}\gg\epsilon_\mathrm{F}$, which allows us to neglect intraband electronic transitions. Above, $v=\sqrt{\epsilon_\mathrm{g}/m}$ is the interband matrix element for the velocity operator, and $\vec{e}=(\vec{e}_\mathrm{x}\pm i \vec{e}_\mathrm{y})/\sqrt{2}$ determines the valley selection by circularly polarized light. We elaborate on the absorption theory in Sec.~\ref{SecXPApsorption}, while the two subsequent sections are devoted to our microscopic theory for XPs. 

\section{Exciton-polaron theory}
\label{sec:SecXPTheory}

\subsection{Exciton problem}
\label{sec:exciton}

Within our model, the exciton can be %well approximated as a composite boson 
described by the creation operator 
\begin{equation}
\label{X}
X_{\vec{p}_\mathrm{X}}^\dagger=\sum_\vec{p} C_{\vec{p}} % \vec{q}} 
e_{\vec{p}+\vec{p}_\mathrm{X}/2}^\dagger h_{-\vec{p}+\vec{p}_\mathrm{X}/2}^\dagger\,.
\end{equation}
Here, $C_{\vec{p}}$ is the wave function for the relative motion of the electron and hole. Note that it does not depend on the exciton's center-of-mass momentum, $\vec{p}_\mathrm{X}$.  This is because there is a separation of relative and center-of-mass dynamics in the two-body problem with a conventional quadratic spectrum. %If the state (\ref{X}) is treated in the variational manner, 
The wave function, $C_\vec{p}$, satisfies the following eigenvalue equation:
\begin{equation}
\label{P}
\big(\epsilon_\vec{p}^\mathrm{e}+\epsilon_{-\vec{p}}^\mathrm{h}\big)C_\vec{p}-\sum_{\vec{p}'} U_{\vec{p}-\vec{p}'}C_{\vec{p}'}=E^\mathrm{X} C_\vec{p}\,,
\end{equation}
with total energy $E^\mathrm{X}$. The resulting spectrum includes both discrete excitonic states and continuous unbound electron-hole pairs, and we label these states by a collective index, $\nu$. The bound excitonic states are labeled by principal $n$ and orbital $l$ quantum numbers, $|\nu\rangle=|n,l\rangle$ (with $|0,0\rangle$ corresponding to the ground-state exciton). 

\subsection{Phenomenological approach}
For completeness, here, we briefly review the theory of exciton-polarons with phenomenological contact exciton-electron interactions~\cite{EfimkinMacDonald1,EfimkinMacDonald2,TMDCDemlerExp}. Its cornerstone is the hierarchy of binding energies for trions ($\epsilon_\mathrm{T}$) and excitons ($\epsilon_\mathrm{X}$), \textit{i.e.},  $\epsilon_\mathrm{T}\ll \epsilon_\mathrm{X}$. This ensures that in the wide density range where $\epsilon_\mathrm{T}\sim \epsilon_\mathrm{F}$, the formation of an exciton by a photo-excited electron and hole is only weakly disturbed by excess charge carriers. In this limit, %we can therefore consider 
it is therefore reasonable to consider the exciton as a structureless quasiparticle where the center-of-mass momentum %$\vec{p}_\mathrm{XP}$ 
is the only relevant degree of freedom.

%Its binding energy $\epsilon_\mathrm{X}\gg \epsilon_\mathrm{T}, \epsilon_\mathrm{F}$ is the largest energy scale which is why an exciton can be considered as a solid structureless quasiparticle with its center of mass momentum $\vec{q}$ to be the only relevant degree of freedom. 

The exciton's interaction with the Fermi sea dresses it into an exciton-polaron. The corresponding Fermi-polaron problem (involving a Fermi sea as a quantum environment) has recently been realized in 2D cold-atom experiments~\cite{Koschorreck2012,Zhang2012,Oppong2019}, and a remarkable understanding of its rich behavior has been achieved~\cite{FermiPolaronReview1,Levinsen2015review}.
%~\cite{FermiPolaron1, FermiPolaron2,FermiPolaron4, FermiPolaron3, FermiPolaronNew3, FermiPolaronDemler,  FermiPolaron5,FermiPolaronNew1,FermiPolaronNew2}. 
The creation operator for an optically active XP with zero momentum %, $\vec{p}_\mathrm{XP}=\vec{0}$, 
can be well approximated by the Chevy ansatz~\cite{FermiPolaron4} as follows: 
\begin{equation}
\label{PF}
P^\dagger_\vec{0}=\phi X_\vec{0}^\dagger + \sum_{\vec{k}\vec{k}'} \chi_{\vec{k} \vec{k}'} X_{\vec{k}'-\vec{k}}^\dagger f_\vec{k}^\dagger f_{\vec{k}'}\,. 
\end{equation}
Above, $\phi$ is the weight of the exciton, while $\chi_{\vec{k} \vec{k}'}$ weights the contribution of a single electron-hole pair excitation of the FS with momenta $k>k_\mathrm{F}$  and $k'<k_\mathrm{F}$. The possibility to excite multiple electron-hole pairs has been proven to be negligibly small~\cite{Combescot2008}. The photo-excited electron and hole interact both with the electron outside the FS and with the hole inside the FS. However, momentum-space limitations mean that the latter is inefficient for $\epsilon_\mathrm{F}\ll\epsilon_\mathrm{X}$ and can therefore be omitted. If we introduce the exciton-electron interactions in a phenomenological manner as $V_\vec{q}$, then the variational equations for the state~\eqref{PF} are given by
%%
%\begin{widetext}
%\begin{equation}
%\begin{split}
%\left (E_\mathrm{X}+\sum_{\vec{k}'} V_\vec{0}\right) \phi + %\sum_{\vec{k}\vec{k}'} V_{\vec{k}-\vec{k}'} \chi_{\vec{k} %\vec{k}'}=E^\mathrm{XP} \phi, \quad\quad 
%(E_\mathrm{X}+ \epsilon_{\vec{k}\vec{k}'}^\mathrm{FS})\chi_{\vec{k}\vec{k'}}+V_{%\vec{k}'- \vec{k}} \phi+\sum_{\bar{\vec{k}}} V_{\bar{\vec{k}}-\vec{k}}  %\chi_{\bar{\vec{k}}\vec{k}'} = E^\mathrm{XP} \chi_{\vec{k}\vec{k'}}.
%\end{split}
%\end{equation}
%\end{widetext}
%%
%
\begin{subequations}
\label{eq:vareqscontact}
\begin{align}
\left(\Delta E^\mathrm{XP}-\sum_{\vec{k}'} V_\vec{0}\right) \phi&= \sum_{\vec{k}\vec{k}'} V_{\vec{k}-\vec{k}'} \chi_{\vec{k} \vec{k}'}\,, \\ %\quad\quad 
\left( \Delta E^\mathrm{XP}- \epsilon_{\vec{k}\vec{k}'}^\mathrm{FS}\right)\chi_{\vec{k}\vec{k'}} & = V_{\vec{k}'- \vec{k}} \phi+\sum_{\bar{\vec{k}}} V_{\bar{\vec{k}}-\vec{k}}  \chi_{\bar{\vec{k}}\vec{k}'}\,.
\end{align}
\end{subequations}
Here, $\epsilon^\mathrm{FS}_{\vec{k}\vec{k}'}=\epsilon^\mathrm{X}_{\vec{k}'-\vec{k}}+\epsilon^\mathrm{f}_\vec{k}-\epsilon^\mathrm{f}_{\vec{k}'}$ is the sum of kinetic energies for the exciton's center of mass and an electron-hole excitation of the Fermi sea. The quantity, $\Delta E^\mathrm{XP}=E^\mathrm{XP}-E^\mathrm{X}$, is the energy of the exciton-polaron state $E^\mathrm{XP}$ defined with respect to the exciton energy $E^\mathrm{X}$. Due to the Pauli-blocking effect of electrons in the FS, the momentum in the last sum of Eq.~(\ref{eq:vareqscontact}) is restricted to $\bar{k}>k_\mathrm{F}$.    

If %the interaction 
%\bar{V}
$V_\vec{q}$ is approximated by a contact potential with a momentum-independent Fourier transform, then the system of equations becomes algebraic~\cite{FermiPolaronNew3} and analytically tractable~\cite{FermiPolaronNew1}. Its solution can be elegantly parameterized by the binding energy $\epsilon_\mathrm{T}$ for the two-particle X-e state, which is a simplified model for the trion. The energy $\epsilon_\mathrm{T}$ determines the splitting between attractive and repulsive XP branches in the limit of vanishing doping, and therefore can be easily adjusted to fit experiments. The frequency and doping dependence of absorption within this model has been extensively discussed in our previous work~\cite{EfimkinMacDonald1} and has been argued to well describe the experimental data. 

\subsection{Microscopic theory}
To derive the microscopic theory for XPs we avoid using excitons as an intermediate step of the theory. Instead we write the creation operator for an XP with zero momentum %, $\vec{p}_\mathrm{XP}=\vec{0}$, 
as follows: 
\begin{equation}
\label{MicroscopicAnzatz}
P_\vec{0}^\dagger=\sum_\vec{p} \phi_\vec{p}e_\vec{p}^\dagger  h_{-\vec{p}}^\dagger + \sum_{\vec{p} \vec{k}\vec{k}'} \chi_{\vec{p}\vec{k} \vec{k}'} e^\dagger_{\vec{p}_+} h^\dagger_{-\vec{p}_-} f_\vec{k}^\dagger f_{\vec{k}'}\,.
\end{equation}
Here, $\vec{p}_\pm=\vec{p}\pm (\vec{k}'-\vec{k})/2$ and $\vec{k}'-\vec{k}$ is the center-of-mass momentum for the photo-excited electron and hole. In addition, $\phi_\vec{p}$ is the wave function for the relative motion of the electron-hole pair, while $\chi_{\vec{p}\vec{k} \vec{k}'}$ weights its correlations with a single particle-hole excitation of the Fermi sea. By neglecting interactions between the excitonic electron or hole and the Fermi-sea hole, since these are inefficient due to momentum-space limitations, we obtain the variational expressions shown below:  
\begin{equation}
\begin{split}
\label{Variational1}
\left(\epsilon^\mathrm{e}_\vec{p}+\epsilon^\mathrm{h}_{-\vec{p}}\right)\phi_\vec{p}- \sum_{\bar{\vec{p}}} U_{\vec{p}-\bar{\vec{p}}} \phi_{\bar{\vec{p}}}\;+ \\ \sum_{\vec{k}\vec{k}'} U_{\vec{k}-\vec{k'}}\left(\chi_{\vec{p}-\frac{\vec{k}-\vec{k}'}{2},\vec{k}\vec{k}'}-\chi_{\vec{p}+\frac{\vec{k}-\vec{k}'}{2},\vec{k}\vec{k}'}\right)= E^\mathrm{XP} \phi_\vec{p}\,,
\end{split}
\end{equation}
and
\begin{equation}
\label{Variational1Part2}
\begin{split}
\left(\epsilon^\mathrm{e}_\vec{p}+\epsilon^\mathrm{h}_{-\vec{p}}+\epsilon_{\vec{k}\vec{k}'}^\mathrm{FS}\right)\chi_{\vec{p}\vec{k}\vec{k}'}-\sum_{\bar{\vec{p}}} U_{\vec{p}-\bar{\vec{p}}} \chi_{\bar{\vec{p}}\vec{k}\vec{k}'}\;+ \\ \sum_{\bar{\vec{k}}} U_{\bar{\vec{k}}-\vec{k}} \left(\chi_{\vec{p}-\frac{\bar{\vec{k}}-\vec{k}}{2},\bar{\vec{k}}\vec{k}'}-\chi_{\vec{p}+\frac{\bar{\vec{k}}-\vec{k}}{2},\bar{\vec{k}}\vec{k}'} \right) +  \\ U_{\vec{k}'-\vec{k}} \left(\phi_{\vec{p}-\frac{\vec{k}'-\vec{k}}{2}}-\phi_{\vec{p}+\frac{\vec{k}'-\vec{k}}{2}} \right) = E^\mathrm{XP} \chi_{\vec{p}\vec{k}\vec{k}'}\,.
\end{split}
\end{equation}
Due to the presence of the Fermi sea, the momenta are restricted to  $k,\,\bar{k}>k_\mathrm{F}$ and $k'<k_\mathrm{F}$ above and in what follows.  
For the considered doping range, $\epsilon_\mathrm{F}\ll\epsilon_\mathrm{X}$, the relative motion of the electron and hole within the exciton is only weakly disturbed by excess electrons. This motivates the decomposition of both wave functions, $\phi_\vec{p}$ and $\chi_{\vec{p}\vec{k}\vec{k}'}$, into the excitonic (and continuum) states discussed in Sec.~\ref{sec:exciton}:
\begin{equation}
\label{Basis}
\phi_\vec{p}=\sum_{\nu} C_\vec{p}^\nu \phi_{\nu}\,,  \quad \quad 
\chi_{\vec{p} \vec{k} \vec{k'}}=\sum_{\nu} C_\vec{p}^\nu \chi_{\nu \vec{k} \vec{k}'}\,.
\end{equation}
The variational equations can then be rewritten as 
%\begin{widetext}
%\begin{equation}
%\label{Variational2}
%E_{\nu}^\mathrm{X} \phi_{\nu}+\sum_{\bar{\nu}\vec{k} %\vec{k'}} \Lambda_{\vec{k}-\vec{k'}}^{\nu \bar{\nu}} %\chi_{\bar{\nu} \vec{k}\vec{k}'}=E^\mathrm{XP} %\phi_{\nu}  \quad \quad \quad \quad 
%(E_\mathrm{X}^\nu+\epsilon_{\vec{k}\vec{k}'}^\mathrm%{FS}) \chi_{\nu \vec{k}\vec{k}'}+\sum_{\bar{\nu}} %\Lambda_{\vec{k}'-\vec{k}}^{\nu \bar{\nu}} %\phi_{\bar{\nu}}  +  \sum_{\bar{\nu} \vec{k}} %\Lambda_{\bar{\vec{k}}-\vec{k}}^{\nu \bar{\nu}} %\chi_{\bar{\nu} \bar{\vec{k}} %\vec{k}'}=E^\mathrm{XP} \chi_{\nu \vec{k}\vec{k}'}  
%\end{equation}
%\end{widetext}
\begin{subequations}
\label{Variational2}
\begin{align}
\Delta E^\mathrm{XP}_\nu \phi_{\nu}&=\sum_{\bar{\nu}\vec{k} \vec{k'}} \Lambda_{\vec{k}-\vec{k'}}^{\nu \bar{\nu}} \chi_{\bar{\nu} \vec{k}\vec{k}'}\,, \\  % \quad \quad \quad \quad 
\left(\Delta E^\mathrm{XP}_\nu -\epsilon_{\vec{k}\vec{k}'}^\mathrm{FS}\right) \chi_{\nu \vec{k}\vec{k}'} &=\sum_{\bar{\nu}} \Lambda_{\vec{k}'-\vec{k}}^{\nu \bar{\nu}} \phi_{\bar{\nu}}  +    \sum_{\bar{\nu} \bar{\vec{k}}} \Lambda_{\bar{\vec{k}}-\vec{k}}^{\nu \bar{\nu}} \chi_{\bar{\nu} \bar{\vec{k}} \vec{k}'}\,,  
\end{align}
\end{subequations}
where we denote $\Delta E^\mathrm{XP}_\nu=E^\mathrm{XP}-E^\mathrm{X}_\nu$. The excitonic levels are coupled by the matrix element $\Lambda^{\nu\bar{\nu}}_\vec{q}$, which represents the amplitude of X-e scattering between states $|i_\nu\rangle=X^\dagger_{\nu \vec{p}_\mathrm{X}} f^\dagger_{\vec{p}_\mathrm{e}}|\mathrm{g}\rangle$ and $|f_\nu\rangle=X^\dagger_{\nu \vec{p}_\mathrm{X}+\vec{q}} f^\dagger_{\vec{p}_\mathrm{e}-\vec{q}}|\mathrm{g}\rangle$ (where $|\mathrm{g}\rangle$ is the undisturbed Fermi sea). The scattering amplitude, with transferred momentum $\vec{q}$, in the Born approximation is given by
\begin{equation}
\Lambda^{\nu\bar{\nu}}_\vec{q}=\langle f_\nu|\big(H_{\mathrm{e\text{-}f}}+H_{\mathrm{h\text{-}f}}\big)|i_{\bar{\nu}} \rangle\,.
\end{equation}
Here, $H_{\mathrm{e\text{-}f}}$ and $H_{\mathrm{h\text{-}f}}$ represent, respectively, the interaction between a Fermi-sea electron and the electron or hole that constitutes the exciton. They correspond to the first two terms in Eq.~\eqref{eq:HC}. The explicit  expression for the scattering matrix element is     
\begin{align}
\begin{split}\
\Lambda^{\nu\bar{\nu}}_{\vec{q}}  =U_\vec{q}\sum_\vec{p} (C_\vec{p}^\mathrm{\nu})^* \left(C_{\vec{p}-\vec{q}/2}^{\bar{\nu}}- C_{\vec{p}+\vec{q}/2}^{\bar{\nu}}\right).
\label{eq:Lambda}
\end{split}
\end{align}

We can further comprehend the nature of $\Lambda^{\nu\bar{\nu}}_\vec{q}$ by now considering its behavior at small momentum transfer, $q a_\mathrm{X}\ll 1$, with $a_\mathrm{X}$ the exciton size. By introducing $C_\vec{r}^\nu$ as the Fourier transform of $C_{\vec p}^\nu$ in Eq.~\eqref{eq:Lambda}, we find that
%At , $\Lambda^{\nu\bar{\nu}}_\vec{q}$ is determined by the matrix element for the dipole moment  $\vec{d}_{\nu \bar{\nu}}=\langle \nu|e\vec{r}|\bar{\nu}\rangle$, $\Lambda^{\nu\bar{\nu}}_\vec{q}=i U_\vec{q} \vec{q}\, \vec{d}_{\nu\bar{\nu}}$, reflecting the charge-dipole nature of the exciton-electron interactions. 
%
\begin{subequations}
\begin{align}
     \Lambda^{\nu\bar{\nu}}_\vec{q}&=  U_\vec{q}\int d\vec{r}\, (C_\vec{r}^\nu)^*C_\vec{r}^{\bar{\nu}}\,2 i \sin\left(\frac{\vec{q}\cdot\vec{r}}{2}\right) \\
     & \simeq i U_\vec{q} \vec{q}\cdot\vec{d}_{\nu\bar{\nu}}\,, \quad qa_X\ll 1\,,
\end{align}
\end{subequations}
where $e\vec{d}_{\nu \bar{\nu}}=\langle \nu|e\vec{r}|\bar{\nu}\rangle$ is the matrix element of the dipole moment. This explicitly reflects the charge-dipole nature of exciton-electron interactions.
%As explicitly demonstrated in App.~\ref{AppMatrixElement}, 
Since $|\nu\rangle$ is a parity eigenstate, $\Lambda^{\nu\bar{\nu}}_\vec{q}$ follows the selection rules for optical dipole transitions between the excitonic states. Therefore, $\Lambda_\vec{q}^{\nu\bar{\nu}}$ is nonzero only between states of opposite parities, which implies that $l-\bar{l}=\pm1,\,\pm3,\,...$\,\,. Consequently, the diagonal matrix elements vanish, $\Lambda^{\nu\nu}_\vec{q}=0$, and we emphasize that only those elements coupling distinct excitonic (and continuum) states are nonzero. 

To proceed, we restrict our attention to the effect of the excess electrons on only the ground excitonic state. The corresponding exciton-polaron state predominantly consists of $\phi \equiv \phi_{\nu} $ and $ \chi_{\vec{k}\vec{k}'} \equiv \chi_{\nu \vec{k}\vec{k}'}$, with $|\nu\rangle=|0,0\rangle$. Provided that $\epsilon_\mathrm{F},\epsilon_\mathrm{T}\ll\epsilon_\mathrm{X}$, the contribution of all excited states is small. Interactions between them are of  second order in $\epsilon_\mathrm{T}/\epsilon_\mathrm{X}$ and can therefore be neglected. If we treat the coupling of excited states to (only) the ground state $|0,0\rangle$ perturbatively, then Eq.~(\ref{Variational2}) simplifies to 
%\begin{widetext}
%\begin{equation}
%\label{Variational3}
%\begin{split}
%\left(E_\mathrm{X}+\sum_{\vec{k}'} V_{\vec{k}' \vec{k}' %\vec{k}'}\right)\phi + \sum_{\vec{k}'} V_{\vec{k} \vec{k}' %\vec{k}'}\chi_{\vec{k}\vec{k}'}=E^\mathrm{XP}\phi, \quad \quad %(E_\mathrm{X}+ \epsilon^\mathrm{FS}_{\vec{k}\vec{k}'})\chi_{\vec{k}\%vec{k'}}+V_{\vec{k}' \vec{k} \vec{k}'} \phi+\sum_{\bar{\vec{k}}} %V_{\bar{\vec{k}} \vec{k} \vec{k}'} \chi_{\bar{\vec{k}}\vec{k}'} = %E^\mathrm{XP} \chi_{\vec{k}\vec{k'}}.
%\end{split}
%\end{equation}
%\end{widetext}
\begin{subequations}
\label{Variational3}
\begin{align}
\left(\Delta E^\mathrm{XP}_0 - \sum_{\vec{k}'} V_{\vec{k}' \vec{k}' \vec{k}'}\right)\phi&= \sum_{\vec k\vec{k}'} V_{\vec{k} \vec{k}' \vec{k}'}\chi_{\vec{k}\vec{k}'}\,, \\ %\quad\quad 
\left(\Delta E^\mathrm{XP}_0 -  \epsilon^\mathrm{FS}_{\vec{k}\vec{k}'}\right)\chi_{\vec{k}\vec{k'}} &=  V_{\vec{k}' \vec{k} \vec{k}'} \phi+\sum_{\bar{\vec k}} V_{\bar{\vec{k}} \vec{k} \vec{k}'} \chi_{\bar{\vec{k}}\vec{k}'}\,.
\end{align}
\end{subequations}
For a detailed derivation we refer the reader to App.~\ref{AppPerturbationTheory}.

This set of equations remarkably coincides with the variational expressions~\eqref{eq:vareqscontact} obtained previously within the phenomenological approach.  However, the microscopically derived interactions, 
\begin{equation}
\label{Vint}
V_{\vec{k}_1 \vec{k}_2 \vec{k}'}=\sum_{\nu \vec{k}}\frac{\Lambda_{\vec{k}-\vec{k}_2}^{0\nu} \Lambda_{\vec{k}_1-\vec{k}}^{\nu0}}{E^\mathrm{X}_0-E^\mathrm{X}_\nu-\epsilon^\mathrm{FS}_{\vec{k} \vec{k}'}}\,,
\end{equation}
are intrinsically nonlocal and depend explicitly on the momentum of the electron before ($\vec{k}_2$) and after ($\vec{k}_1$)  scattering with the exciton, as well as on the momentum of the redundant Fermi hole ($\vec{k}'$) which does not participate directly in the scattering process.  Note that we have taken $E^\mathrm{XP}\simeq E_0^\mathrm{X}$ in the denominator of Eq.~(\ref{Vint}).  This approximation makes Eq.~(\ref{Variational3}) tractable, and is reasonable because the total energy is only shifted away from the exciton energy by a small amount (on the order of $\epsilon_\mathrm{F},\epsilon_\mathrm{T}\ll\epsilon_\mathrm{X}$).

%The cornerstone relation of the exciton-polaron theory is the hierarchy of scales $\epsilon_\mathrm{F}\sim \epsilon_\mathrm{T} \ll \epsilon_\mathrm{X}$ that makes its applicable in the wide density range for excess charge carriers. At very low doping $\epsilon_\mathrm{F}\ll \epsilon_\mathrm{T}$, three-particle correlations not captured by the theory can be expected to be important. However, predictions of trion and exciton-polaron scenarios have been recently found to be almost indistinguishable in this regime that further extends the applicability of the developed microscopic exciton-polaron theory.}
The cornerstone relation of the exciton-polaron theory is the hierarchy of scales $\epsilon_\mathrm{F}\sim \epsilon_\mathrm{T} \ll \epsilon_\mathrm{X}$, ensuring that the theory is applicable in a wide range of density of excess charge carriers. At very low doping, $\epsilon_\mathrm{F}\ll \epsilon_\mathrm{T}$, three-particle correlations not captured by the theory may be expected to be important. However, predictions of trion and exciton-polaron scenarios have been recently found to be almost indistinguishable in this regime~\cite{GalzovXT}, thus suggesting that the theory is applicable also at low doping.

%Thus far, 
Note that our arguments thus far do not specify the interaction between electrons and holes, $U_\vec{q}$, which relies on details of the screening (dielectric or metallic) due to the surrounding media. For this reason, our developed theory is applicable to monolayer semiconductors as well as to conventional semiconductor quantum wells (QWs). In the following, for simplicity, we have chosen to use the 2D Coulomb potential, $U_\vec{q}=2\pi e^2/\kappa q$, where $\kappa$ is the effective dielectric constant of the surrounding environment. While this potential properly captures only the excited excitonic states including the continuous part of the spectrum, analytic expressions for the wave functions $C^\nu_\vec{p}$ are known, which considerably simplifies the numerics.  These wave functions are provided in App.~\ref{AppExcitonicSpectrum}.  (Note, the entire spectrum is reasonably well described by the Keldysh potential~\cite{Keldysh1,Keldysh2,Keldysh3} which properly incorporates the dielectric screening in layered systems~\cite{TMDCEx1,TrionExperiment4}.)  In the discussion, Sec.~\ref{SecConc}, we argue that our results are very general and therefore extendable to other potentials.   

\section{Exciton-electron interactions}
\label{Sec:X-e}
The excitonic spectrum with 2D Coulomb interactions, $U_\vec{q}=2\pi e^2/\kappa q$,  admits an analytical solution which has been extensively discussed in Refs.~\cite{Coulomb2D1,Coulomb2D2,Coulomb2DPortnoi} and is summarized in App.~\ref{AppExcitonicSpectrum}. The spectrum of discrete states, $E^\mathrm{X}_\nu=-\epsilon_\mathrm{X}/(2n+1)^2$ where $|\nu\rangle=|n,l\rangle$, possesses accidental degeneracy which is typical for Coulomb problems~\cite{Coulomb2DPortnoi}. Here, $\epsilon_\mathrm{X}=\hbar^2/ma_\mathrm{X}^2=me^4/\hbar^2\kappa^2$ is the binding energy of the excitonic ground state $|\nu\rangle=|0,0\rangle$, and $a_\mathrm{X}=\hbar^2 \kappa/m e^2$ is the average electron-hole separation in that state. The continuous spectrum of unbound electron-hole pairs, $E^\mathrm{X}_\nu=q^2\epsilon_\mathrm{X}$, can be labeled $|\nu\rangle=|q,l\rangle$ by a dimensionless absolute momentum $q$ and an orbital quantum number $l$. For numerical calculations of the effective X-e interactions, we have taken into account excitonic states with $n=1,\,...,\,10$ and $|l|=1,\,3,\,5$ as well as the whole continuous spectrum with $|l|=1,\,3$. We have checked that the contribution of truncated states is negligibly small. 

The derived X-e interaction~(\ref{Vint}) is doping dependent and nonlocal in nature.  However, it is instructive to introduce its simplified doping-dependent \emph{local} counterpart:
\begin{equation}
\label{Vintlocal}
V_\mathrm{L}(\vec{q})=V_{\vec{q},\vec{0},\vec{0}}=V_{\vec{0},\vec{q},\vec{0}}\,.    
\end{equation}
This local potential is solely determined by the transferred momentum $\vec{q}$, which implies that the corresponding real-space potential $V_\mathrm{L}(\vec{r})$ only depends on the relative distance between the exciton and electron.     

Due to the hierarchy of energy scales $\epsilon_\mathrm{F},\,\epsilon_\mathrm{T}\ll\epsilon_\mathrm{X}$ the spatial scale over which the polaronic correlations occur can be estimated by $a_\mathrm{T}\approx a_\mathrm{X} \sqrt{\epsilon_\mathrm{X}/\epsilon_\mathrm{T}}\approx 3 a_\mathrm{X}$. This implies that the details of the real-space potential profile, corresponding to  Eq.~(\ref{Vint}), are most important at $r\gtrsim a_\mathrm{X}$. A careful numerical comparison of the nonlocal~(\ref{Vint}) and local~(\ref{Vintlocal}) potentials is presented in App.~\ref{AppNonlocal}, and there we show how the latter very well approximates the interactions for $q a_\mathrm{X}\ll 1$. Evidently, the nonlocal nature of X-e interactions is unimportant, and at $r \gtrsim a_\mathrm{X}$ they are well captured by Eq.~(\ref{Vintlocal}) which we use below. 

\begin{figure}[t]
	\includegraphics[width=8cm ]{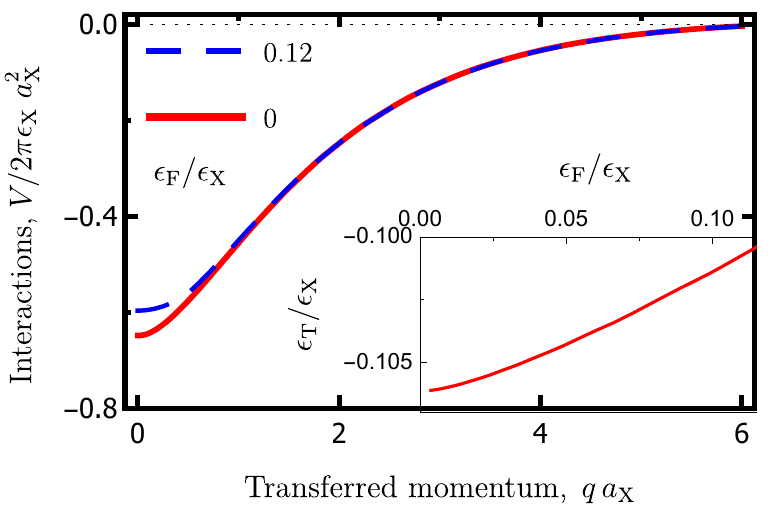}
	\caption{Fourier transform of the X-e interactions at doping levels $\epsilon_\mathrm{F}=0$ [solid red] and $\epsilon_\mathrm{F}=0.12\,\epsilon_\mathrm{X}$ [dashed blue]. The inset shows the weak doping dependence for the binding energy $\epsilon_\mathrm{T}$ of the two-particle X-e bound state. \label{FigVFourier}}
	%\vspace{1cm}
\end{figure}

The resulting local interactions, $V_\mathrm{L}(\vec{q})$, at zero doping and at doping $\epsilon_\mathrm{F}=0.12\,\epsilon_\mathrm{X}$ are presented in Fig.~\ref{FigVFourier}. %For the latter case, as clearly seen in Fig.~\ref{FigOpticalConductivity}, the repulsive exciton-polaron peak is almost depleted and only the attractive peak is present. 
The doping dependence of the interaction is weak and its dependence on momentum at $q a_\mathrm{X}\ll 1$ is smooth. This is a signature of the short-range nature of the interactions and it presents an additional justification for the contact phenomenological potential that has been used previously~\cite{EfimkinMacDonald1,EfimkinMacDonald2,TMDCDemlerExp}. 

In the limit of vanishing doping, the polaronic physics reduces to the two-particle X-e problem. Their bound state represents a simplified model for the trion and its binding energy is shown in the inset of Fig.~\ref{FigVFourier}. It  has a very weak doping dependence and is equal to $\epsilon_\mathrm{T}^\star=0.106\,\epsilon_\mathrm{X}$ at $\epsilon_\mathrm{F}=0$. This energy is quite close to the numerical solution for the three-particle trion problem, $\epsilon_\mathrm{T}=0.118\,\epsilon_\mathrm{X}$, with 2D Coulomb interactions~\cite{TrionEnergy1,TrionEnergy2,TrionEnergyRecent}. This suggests that the low-doping regime $\epsilon_\mathrm{F}\ll \epsilon_\mathrm{T}$ is also properly captured by the polaronic theory, in agreement with the conclusions of Ref.~\cite{GalzovXT} where a detailed comparison of the predictions of trion and XP theories is presented. The origin of the small discrepancy with the numerical solution for $\epsilon_\mathrm{T}$ is the perturbative treatment of the excited exciton states --- not the local approximation of interactions given by Eq.~(\ref{Vintlocal}).

\begin{figure}[t]
	\includegraphics[width=8 cm]{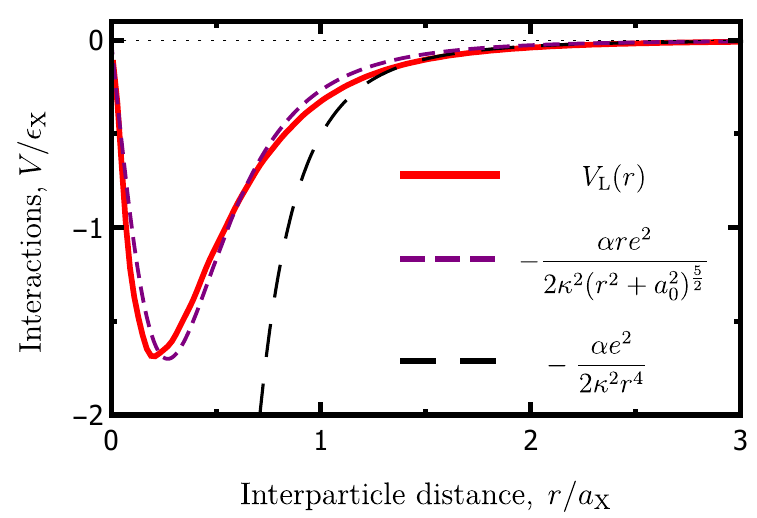}
	\caption{Spatial profile of the local X-e interaction $V_\mathrm{L}(\vec{r})$ [solid red], its approximation %fitting 
	by a modified charge-dipole potential $V_\mathrm{A}(\vec{r})$ [dashed purple], and the asymptotic behavior at large interparticle distance [dashed black]. \label{FigVReal}
	}
\end{figure}

The spatial dependence of the interactions at $\epsilon_\mathrm{F}=0$ is presented in Fig.~\ref{FigVReal}, and it can be accurately approximated by the following expression: 
\begin{equation}
\label{VintAnalytic}
V_\mathrm{A}(\vec{r})=\frac{\alpha e^2 r}{2 \kappa^2 (r^2+a_0^2)^\frac{5}{2}}\,.
\end{equation}
Here, $a_0\approx0.54\,a_\mathrm{X}$, and $\alpha$ is the polarizability of the ground  state $|0,0\rangle$ which, within second-order perturbation theory, is calculated to be
\begin{equation}
\alpha=\sum_{\nu}\frac{|\vec{d}_{0\nu}|^2}{E_0^\mathrm{X}-E^\mathrm{X}_{\nu}}\,.
\end{equation}
As before, $\vec{d}_{0\nu}$ is the matrix element for the dipole moment between the excitonic ground state and an excited state $|\nu\rangle$. It is clearly seen in Fig.~\ref{FigVReal} that the  potential perfectly follows the classical charge-dipole interaction, %$V(\vec{r})=
$-\alpha e^2/2 \kappa^2 r^4$ at $r\gtrsim a_\mathrm{X}$, which is the relevant region for polaronic physics. For this reason, we refer to the microscopically derived interaction~(\ref{Vint}) as a charge-dipole potential. 

Numerically, we find that the contributions of discrete excitonic states ($\alpha_\mathrm{D}\approx2\alpha/3$) and unbound electron-hole pairs ($\alpha_\mathrm{C}\approx\alpha/3$) to the polarizability of the exciton are comparable with each other. Thus, the latter is important and cannot be truncated within the microscopic XP theory. It should also be mentioned that the low-energy continuum states cannot be approximated by plane waves, due to %which can be interpreted as 
the presence of Gamow-Sommerfeld enhancement~\cite{Glutch}.   

\begin{figure}[t]
	\includegraphics[width=8 cm]{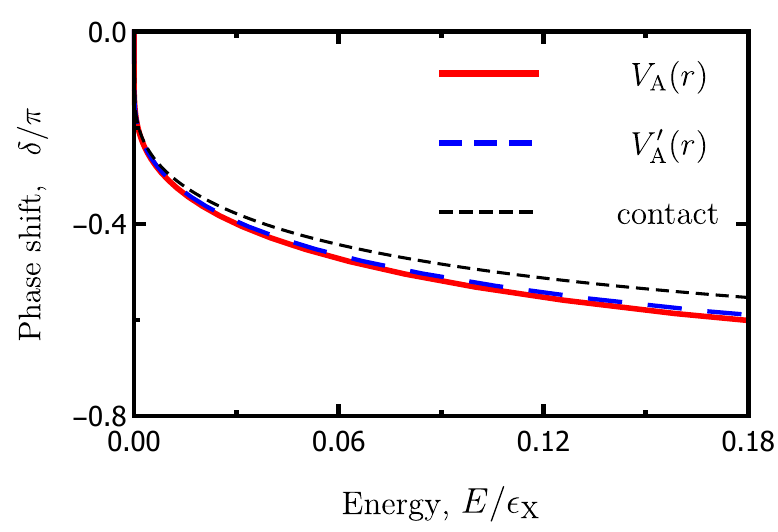}
	\caption{Energy dependence of the X-e scattering phase shift $\delta(E)$ with the interactions approximated by $V_\mathrm{A}(\vec{r})$ [solid red] and $V_\mathrm{A}'(\vec{r})$ [long-dashed blue]. The result for the contact potential which reproduces the X-e binding energy $\epsilon_\mathrm{T}^\star$ is also shown for comparison [short-dashed black]. For all considered interactions, the phase shift approaches $\delta=-\pi$ for $E\gg\epsilon_\mathrm{X}$.  \label{FigPhaseShift}
	}
\end{figure}

%X-e 
The interactions between excitons and electrons have also been recently addressed through other approaches. In particular, a comprehensive  analysis~\cite{ImamogluSchmidt}  of  the  energy-dependent  phase  shifts $\delta(E)$ for X-e scattering has demonstrated that their low-energy  behavior  is  well captured by the following modified charge-dipole interaction:
\begin{equation}
V'_\mathrm{A}(\vec{r})= 
\begin{cases}
-\frac{\alpha}{2}\left(\frac{\partial U}{\partial r}\right)^2, \quad & r>a_*\,, \\    
\quad \quad V_*\,, \quad \quad \; & r<a_*\,,
\end{cases}
\end{equation}
which importantly features the same functional form of the long-range interactions.
Here, the potential $V_*$ and length $a_*$ (which is comparable to the exciton radius $a_\mathrm{X}$) are parameters used for the fitting of $\delta(E)$ calculated numerically within the three-particle problem. Within the spatial range relevant for the polaronic correlations, $r\gtrsim a_\mathrm{T}\sim 3 a_\mathrm{X}$, the potential $V'_\mathrm{A}$ follows the classical charge-dipole behavior in the same manner as $V_\mathrm{A}(\vec{r})$ in Eq.~(\ref{VintAnalytic}). However, they have different short-range behavior at $r\lesssim a_\mathrm{X}$. In Ref.~\cite{ImamogluSchmidt}, the interactions $U(\vec{r})$ were approximated by the Keldysh potential, which complicates a direct comparison between $V_\mathrm{A}$ and $V'_\mathrm{A}$, however we may still compare the general behavior of the resulting X-e scattering. To investigate the importance of short-range details for the X-e potentials, we therefore instead use 2D Coulomb interactions, $U(\vec{r})=e^2/\kappa r$, and we choose the potential $V'_\mathrm{A}$ to be continuous, which implies that $V_*=-\alpha e^2/2\kappa^2 a_*^4$. We then adjust the length $a_*=0.866\,a_\mathrm{X}$ such that the potential $V'_\mathrm{A}$ reproduces the binding energy of the exciton-electron bound state $\epsilon_\mathrm{T}^\star$ in this simplified model for the formation of a trion. The resulting energy dependence of the scattering phase shift $\delta(E)$ is presented in Fig.~\ref{FigPhaseShift} where we see that the results for $V_\mathrm{A}$ and $V'_\mathrm{A}$ are almost indistinguishable, demonstrating that the short-range details of the potentials are not important. We can further illustrate this point by comparing with the universal form of the phase shift calculated from a contact exciton-electron potential that reproduces the same binding energy, $\cot\delta(E)=\frac1\pi\ln(E/\epsilon_\mathrm{T}^\star)$. We see that the phase shifts obtained within the potentials $V_\mathrm{A}$ and $V'_\mathrm{A}$ can be reasonably described by contact interactions, which does not agree with the conclusions of Ref.~\cite{ImamogluSchmidt}. Our calculations therefore suggest that the screening of electronic interactions is essential to resolve this discrepancy.            

%and for the contact potential chosen to reproduce the X-e binding energy $\epsilon_\mathrm{T}^*$. As is clearly seen in Fig.~(\ref{FigPhaseShift}) It should be mentioned that the calculated energy dependence of the phase shift $\delta(E)$ can be reasonably described by contact interactions, which does not agree with the conclusions of Ref.~\cite{ImamogluSchmidt}. Our calculations suggest that the screening of interactions is essential to resolve this discrepancy.   

Previous work has argued that the X-e interactions can be extracted directly from the matrix element of their scattering, $V_\mathrm{S}(\vec{q})=\Lambda^{00}_\vec{q}$~\cite{XEint1,XPnew1, XEint2}. Its magnitude and sign were shown to be very sensitive to the ratio between electron $m_\mathrm{e}$ and hole $m_\mathrm{h}$ masses. In particular, for the balanced case considered here where $m_\mathrm{e}=m_\mathrm{h}$, $V_\mathrm{S}(\vec{q})$ is zero due to parity arguments and it is therefore insensitive to the details of the interaction potential, $U_\vec{q}$, between electrons and holes~\footnote{In the balanced case $m_\mathrm{e}=m_\mathrm{h}$, the nonzero $V_\mathrm{S}(\vec{q})$ was mistakenly claimed in Ref.~\cite{XPnew1} but resolved thereafter~\cite{XEint2}. Private communication with David Reichman}. %The vanishing of the matrix element, $V_\mathrm{S}$, reflects the charge-dipole nature of the derived X-e interactions. 
Therefore, our theory represents the minimal level of perturbation theory necessary to describe the experimentally relevant case of TMDC monolayers.
%We also note that $V_\mathrm{S}(\vec{q})$ is nonzero for the case of mass imbalance and is dominant for the strong imbalance present in the case of semiconductor QWs ($m_\mathrm{e}/m_\mathrm{h}$ is $0.3$ for $\hbox{GaAs}$ and $0.45$ for $\hbox{CdTe}$)}. 

The matrix element $V_\mathrm{S}(\vec{q})$ can be %also 
nonzero in the presence of the exchange channel that appears if the electron within the exciton and those of the Fermi sea are indistinguishable. This is relevant in the case of intravalley exciton-electron correlations in TMDC monolayers and spin-triplet ones in semiconductor QWs, however this effect is outside the scope of the present work.
%In another work~\cite{XPnew1}, X-e interactions were suggested to be extractable from the matrix element of their scattering, $V_\mathrm{S}(\vec{q})=\Lambda^{00}_\vec{q}$. However, we note that this is zero, $\Lambda^{00}_\vec{q}=0$, due to parity arguments (also see App.~\ref{AppMatrixElement}), and it is therefore insensitive to the details of the interaction potential, $U_\vec{q}$, between electrons and holes. The nonzero $V_\mathrm{S}(\vec{q})$ that was claimed in Ref.~\cite{XPnew1} originates from a mistake that is briefly discussed in App.~\ref{AppXescattering}. The vanishing of the scattering amplitude, $V_\mathrm{S}$, reflects the fact that X-e interactions are due to the polarization of the exciton by the electric field from the electron. This process appears at second order in perturbation theory, with no contribution at first order. 

We conclude that the interactions between an exciton and an electron of the Fermi sea, Eq.~(\ref{Vint}), are very well approximated by the local and doping independent interactions in Eq.~(\ref{VintAnalytic}) that are derived from the two-particle exciton-electron problem. We use the latter in our numerical calculations of absorption by XPs.

%For the ensuing numerical calculations of absorption by XPs, we use the modified  doping-independent interaction potential given by Eq.~(\ref{VintAnalytic}).

%For the ensuing numerical calculations of absorption by XPs, we use the local doping-dependent interaction potential given by Eq.~(\ref{Vintlocal}).

\section{Absorption of exciton-polarons}
\label{SecXPApsorption}
\subsection{Derivation of optical conductivity}

The absorption of a semiconductor is determined by the real part of its optical conductivity $\sigma(\omega)$. Within linear-response (Kubo) theory, this can be approximated by~\footnote{The optical conductivity is an even function of the frequency $\omega$, and within linear response theory, can be presented as $\sigma_\mathrm{total}(\omega)=\sigma(\omega)+\sigma(-\omega)$, with $\sigma(\omega)$ given by Eq.~(\ref{OpticalConductivity}). In the frequency range $\omega\sim\epsilon_\mathrm{g}$, the second term is negligible and can be safely omitted.}
\begin{equation}
\label{OpticalConductivity}
\sigma(\omega)=\frac{\pi}{2 \epsilon_\mathrm{g}}\sum_\mu  |\langle \mathrm{g}|{\vec{J}} | \mu \rangle|^2 \delta\big(\hbar \omega-E^\mathrm{XP}_\mu\big)\,. 
\end{equation}
Above, ${\vec{J}}= e v \vec{e} \sum_\vec{p} e_\vec{p} h_{-\vec{p}} + \mathrm{h.c.}$ is the electric current operator restricted to intraband transitions. Its matrix element is calculated between the ground state, $|g\rangle=\Pi_{k'<k_\mathrm{F}} f^\dagger_{\vec{k}'} |\mathrm{vac}\rangle$, and the polaronic state with zero momentum, $|\mu\rangle=P_{\mu\vec{0}}^\dagger|g\rangle$. The index $\mu$ labels an eigenstate of the polaronic equations~(\ref{Variational3}) with  energy $E^\mathrm{XP}_\mu$.  The corresponding matrix element is given by 
\begin{equation}\langle \mathrm{g}|{\vec{J}} |\mu \rangle=e v \vec{e} \sum_\vec{p} \phi_{\mu \vec{p}}= e v \vec{e} \sum_\nu C^\nu_{\vec{r}\,=\,0} \phi_{\mu \nu}\,. 
\end{equation}
Here, we have used the decomposition $\phi_{\mu\vec{p}}=\sum_\nu C^\nu_\vec{p} \phi_{\mu\nu}$ in terms of the excitonic states $C_\vec{p}^\nu$, as written in~(\ref{Basis}). The contribution of excited states is small at $\epsilon_\mathrm{F}\ll\epsilon_\mathrm{X}$ and we only keep the contribution of the ground state $\phi_\mu\equiv \phi_{\mu 0}$. If we assume that the light is circularly polarized and note that $C^0_{\vec{r}\,=\,0}=\sqrt{2/\pi a_\mathrm{X}^2}$, then the optical conductivity~(\ref{OpticalConductivity}) simplifies as follows:
\begin{equation}
\sigma(\omega)=\sigma_0 \epsilon_\mathrm{X}\sum_\mu 2\pi |\phi_{\mu}|^2  \delta (\omega-E_\mu^\mathrm{XP})=\sigma_0 \epsilon_\mathrm{X} A_\mathrm{X}(\omega)\,.
\end{equation}
Above, $\sigma_0=e^2/2\pi \hbar$ is the conductivity quantum and we have introduced the spectral function for excitons dressed into XPs, $A_\mathrm{X}(\omega)$. The latter naturally appears in the diagrammatic theory for XPs~\cite{EfimkinMacDonald1,EfimkinMacDonald2,TMDCDemlerExp}. The frequency dependence of the absorption is completely determined from the spectral function, and we have numerically calculated this by using the variational equations~(\ref{Variational3}) for XPs with local doping-dependent interactions~(\ref{Vintlocal}).

%Its frequency dependence solely determines the absorption and it has been numerically calculated based on the variational equations for XPs, Eq.~(\ref{Variational3}), with local doping dependent interactions, Eq.~(\ref{Vintlocal}).   

\subsection{Doping dependence of optical conductivity}
We now proceed to compare the predictions of our parameter-free microscopic theory for XPs with those of the phenomenological approach. To this end, we need to adjust $\epsilon_\mathrm{T}$ in the latter, since this is the only free parameter. We have chosen to use $\epsilon_\mathrm{T}^\star$ which represents the binding energy for the two-body X-e state. The numerical calculations require a broadening~\cite{Parish2016}, and thus we replace $\delta(\omega)\to (\gamma_\mathrm{X}/\pi)/(\omega^2+\gamma_\mathrm{X}^2)$, %$\omega\rightarrow \omega+i\gamma_\mathrm{X}$, and we have used
where we take
$\gamma_\mathrm{X}=0.1\,\epsilon_\mathrm{T}^\star$. Physically, such a broadening can originate from the radiative decay of excitons or from their scattering due to disorder or phonons; all of these effects are inevitably present in real materials. 

\begin{figure}[b]
	\includegraphics[width=8 cm]{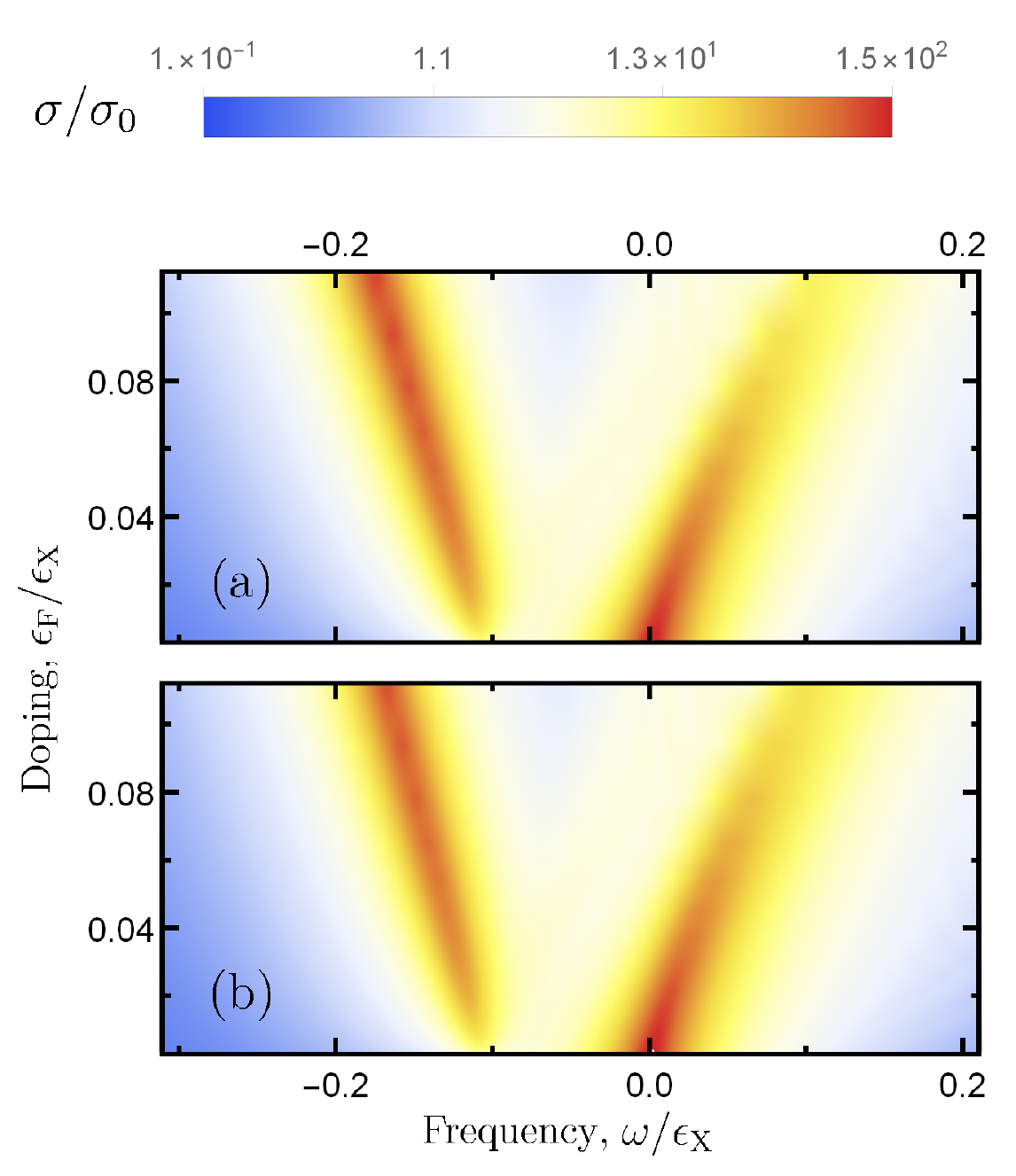}
	\caption{Optical conductivity, $\sigma(\omega)/\sigma_0$, where $\sigma_0 =e^2/h$ is the quantum unit of conductance. Interactions between an exciton and a Fermi sea of excess charge carriers are approximated by either a charge-dipole (a) or a contact (b) potential, which causes the excitonic absorption feature to split into attractive and repulsive exciton-polaron branches.  \label{FigOpticalConductivity}}
\end{figure}

The doping and frequency dependence of the optical conductivity $\sigma(\omega)$ is presented in Fig.~\ref{FigOpticalConductivity}, both for our charge-dipole interactions and for the case of contact interactions~\cite{Parish2016,EfimkinMacDonald1}. The dressing of excitons by the Fermi sea of excess charge carriers splits them into attractive (redshifted peak) and repulsive (blueshifted) XPs. Even at a quantitative level, the predictions between contact and charge-dipole interactions agree very well, and any difference becomes apparent only at a moderate doping $\epsilon_\mathrm{F}\sim\epsilon_\mathrm{T}$.  In the following, we characterize these attractive (A) and repulsive (R) branches 
by their position $\bar{\omega}_{\mathrm{A}(\mathrm{R})}$, width $\gamma_{\mathrm{A}(\mathrm{R})}$, and spectral weight $Z_{\mathrm{A}(\mathrm{R})}$ (also commonly referred to as the oscillator strength).  

%At a qualitative level, they look similar and the dressing of excitons by the Fermi sea of excess charge carriers splits them into attractive (redshifted peak) and repulsive (blueshifted) XPs. However, it is clearly seen that the finite range of the interactions strongly impacts on the spectral weight redistribution between the XP branches. In the following, we characterize these attractive (A) and repulsive (R) branches %can be characterized 
%by their position $\bar{\omega}_{\mathrm{A}(\mathrm{R})}$, width $\gamma_{\mathrm{A}(\mathrm{R})}$, and spectral weight $Z_{\mathrm{A}(\mathrm{R})}$ (also commonly referred to as the oscillator strength).    

\begin{figure}
    \includegraphics[width=8 cm]{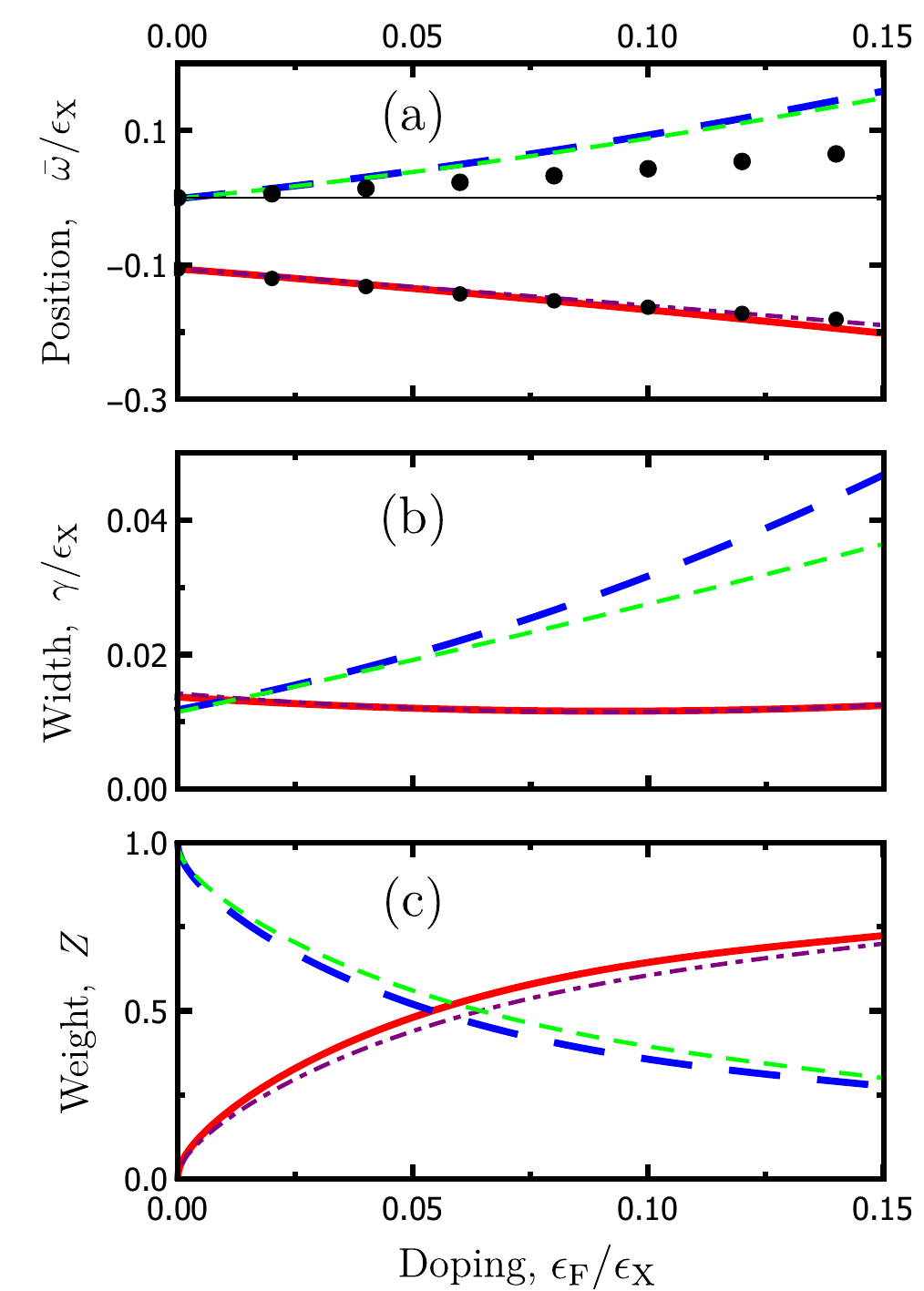}
	\caption{Doping dependence of the peak position (a), peak width (b), and the associated spectral weight (c). The solid red and long-dashed blue lines correspond, respectively, to the attractive and repulsive XP branches calculated by using charge-dipole interactions. The dot-dashed purple and short-dashed green lines likewise correspond to those calculated for contact interactions. Black dots in (a) correspond to the model with infinite exciton mass and are evaluated with Eqs.~(\ref{FummiTheorem})} \label{FigPeaks} 
\end{figure}

\begin{figure}
	\includegraphics[width=8 cm]{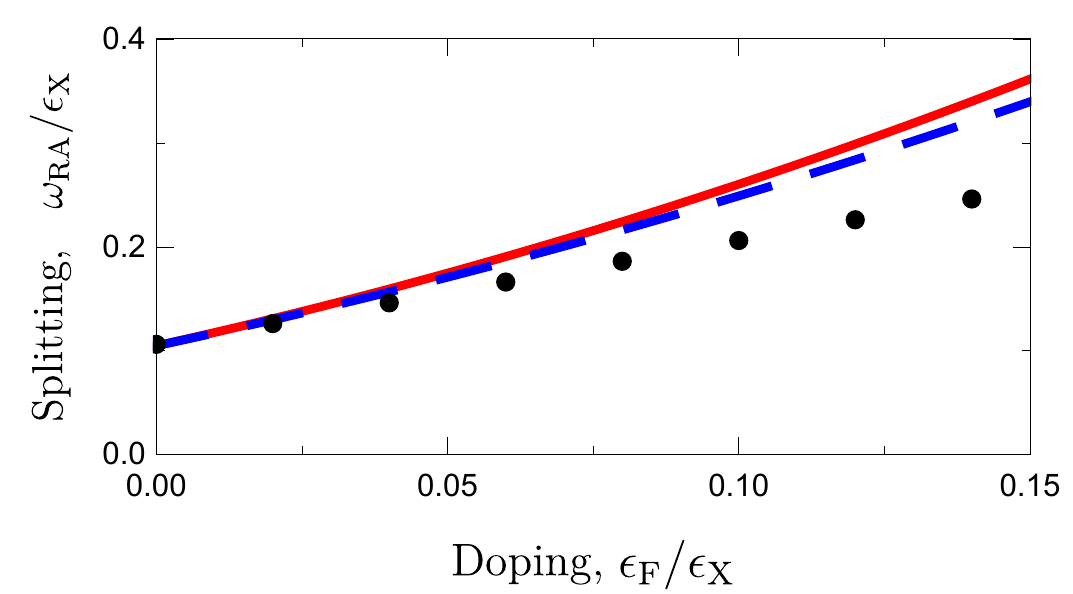}
	\caption{ Doping dependence of the splitting between the attractive and repulsive XP branches for both charge-dipole [solid red] and contact [dashed blue] interactions. It is well captured by relation $\Delta\bar{\omega}=\epsilon_\mathrm{T}+3 \epsilon_\mathrm{F}/2$. Black dots in (a) correspond to the model with infinite exciton mass that predicts $\Delta\bar{\omega}=\epsilon_\mathrm{T}+\epsilon_\mathrm{F}$.}  \label{FigSplitting}
\end{figure}

The doping dependence of the peak positions $\bar{\omega}_{\mathrm{A}(\mathrm{R})}$ is presented in Fig.~\ref{FigPeaks}(a). It is important to separate the effect of excess charge carriers on the splitting between XPs ($\Delta \bar{\omega} = \bar{\omega}_\mathrm{R}-\bar{\omega}_\mathrm{A}$) and their synchronous shift. The latter can also be induced by other factors (\textit{e.g.}, band gap renormalization and interactions with phonons) that are material dependent and are not easy to separate and distinguish. By contrast, the splitting %$\omega_{\mathrm{R}\mathrm{A}}=\omega_\mathrm{R}-\omega_\mathrm{A}$
originates solely from the polaronic dressing and is presented in Fig.~\ref{FigSplitting}. The difference between the predictions of the two models becomes apparent only at $\epsilon_\mathrm{F}\approx\epsilon_\mathrm{T}$, but is still much smaller than $\Delta \bar{\omega}$. Importantly, within both models, the splitting increases linearly with $\epsilon_\mathrm{F}$ as $\Delta\bar{\omega}=\epsilon_\mathrm{T}+3 \epsilon_\mathrm{F}/2$, which can be considered as a hallmark of Fermi polaron physics and agrees with experimental observations~\cite{TMDC5,TrionExperiment1}. The factor $3/2$ is given by the ratio between the electron mass $m$ and the reduced exciton-electron mass $2m/3$. 

It is instructive to compare these results with the predictions of the exciton-polaron model where the exciton is taken to be infinitely heavy. This scenario is known to be exactly solvable, and according to Fumi's theorem the position of the peaks are related to the energy dependent phase shifts $\delta(E)$ for  X-e  scattering as follows~\cite{Fumi1,Fumi2}
\begin{equation}
\label{FummiTheorem}
\begin{split}
\omega_\mathrm{R}=-\int_0^{\epsilon_\mathrm{F}}d E \frac{\delta(E)}{\pi}  \\ 
\omega_\mathrm{A}=-\epsilon_\mathrm{T}-\epsilon_\mathrm{F}-\int_0^{\epsilon_\mathrm{F}}d E \frac{ \delta(E)}{\pi}     
\end{split}
\end{equation}
The corresponding results are also shown in Figs.~\ref{FigPeaks}(a) and ~\ref{FigSplitting}. The position of attractive branch is well captured by the infinite mass model, while the shift of the repulsive one is underestimated. The doping dependence of the splitting between peaks is also linear, $\Delta \bar{\omega}=\epsilon_\mathrm{T}+\epsilon_\mathrm{F}$, but with the factor $1$ instead of $3/2$. This is not surprising, since the reduced exciton-electron mass equals the electron mass $m$ in the case of the infinitely heavy exciton.

The doping dependence of the peak width $\gamma_\mathrm{A(R)}$ is presented in Fig.~\ref{FigPeaks}(b). In the limit of vanishing doping, $\epsilon_\mathrm{F}\ll \epsilon_\mathrm{X}$, the widths of both XP branches are equal to the bare broadening for excitons $\gamma_\mathrm{X}$. At finite doping, the repulsive XP linearly broadens with $\epsilon_\mathrm{F}$, while the width of the attractive XP remains unchanged. This behavior agrees with observations both in TMDC monolayers~\cite{TrionExperiment1} and in semiconductor QWs~\cite{QW1,QW5}, and it is produced by both models with only a small difference between them, even at moderate doping, $\epsilon_\mathrm{F}\sim\epsilon_\mathrm{T}$. 

Figure \ref{FigPeaks}(c) shows the doping dependence of the spectral weight $Z_{\mathrm{A}(\mathrm{R})}$. While the total weight is conserved, $Z_\mathrm{R}+Z_\mathrm{A}=1$, the doping results in a flow from the repulsive XP branch to the attractive XP branch until the former disappears. 
%The minor discrepancy from the expected values of $Z_\mathrm{A}=0$ and $Z_\mathrm{R}=1$ at $\epsilon_\mathrm{F}=0$ is an artifact of our numerical calculations: The presence of the bare broadening $\gamma_\mathrm{X}$ results in the absorption features merging due to XPs at very low doping, $\epsilon_\mathrm{F}\lesssim \gamma_\mathrm{X}$, and this prevents us from distinguishing them clearly. We see that the model with contact interactions slightly underestimates the spectral flow rate, and the difference between the predictions is clearly seen in the wide density range. 
Again, we see a very good agreement between the results of the charge-dipole potential and those from the contact exciton-electron interaction.
At a qualitative level, the behavior of the weights agrees with the experiment~\cite{TrionExperiment1}, which reported the dependence of absorption on the gate voltage that electrically controls the charge-carrier concentration in a TMDC monolayer.

%and the repulsive branch survives the doping up to $\epsilon_\mathrm{F}\sim 2 \epsilon_\mathrm{T}$~\cite{EfimkinMacDonald1}.  By contrast, the model with charge-dipole interactions predicts that the repulsive XP branch disappears at a doping of $\epsilon_\mathrm{F}\sim \epsilon_\mathrm{T}$. At a qualitative level, this behavior agrees with experiments~\cite{TrionExperiment1} which have reported the dependence of the absorption on the gate voltage~\cite{TrionExperiment1} that electrically controls the charge-carrier concentrations in a TMDC monolayer. However, the dependence of the absorption on the excess charge-carrier concentration is needed to confirm the effect of the finite range of X-e interactions in absorption. 

\section{Discussion}
\label{SecConc}
In this work, we have focused on the experimentally relevant scenario where the photo-excited electron is distinguishable from those of the Fermi sea. The corresponding band arrangement, presented in Fig.~\ref{FigBands}, implies the presence of valley splitting in order for one of the Fermi seas to be depleted. 
Since the two valleys are time reversal partners, this splitting 
can be induced either by exchange coupling with a ferromagnetic substrate or by a magnetic field~\cite{TMDCMF1,TMDCMF2,TMDCMF3,TMDCMF4}. As shown in previous work by some of us~\cite{EfimkinMacDonald2}, the magnetic field only weakly influences the polaronic physics as long as $\hbar \omega_\mathrm{B}\ll\epsilon_\mathrm{T}$, with $\omega_\mathrm{B}=eB/mc$ the Larmor frequency for electrons. For a TMDC monolayer with $\epsilon_\mathrm{T}\approx 20\; \mathrm{meV}$, this condition corresponds to $B\ll 57 \; \mathrm{T}$, which is why even a magnetic field  $B\sim 20~\mathrm{T}$ can still be considered rather weak. The combined effect of a magnetic substrate $\mathrm{EuS}$~\cite{MagneticSubstarte} and a magnetic field $23~\mathrm{T}$~\cite{MagneticField} can be estimated to be $7.5\;\mathrm{meV}$ which is sufficient to probe solely intervalley polaronic correlations, free from Pauli blocking and exchange physics.

We have demonstrated that at $\epsilon_\mathrm{F}\ll\epsilon_\mathrm{X}$, the X-e interactions can be approximated by a local potential that depends only on the relative distance between the exciton and the electron, and that its dependence on the doping is very weak. As a result, the presence of a Fermi sea of excess charge carriers only weakly influences the interactions, and they can be effectively addressed from the perspective of the three-particle trion problem. The recent comprehensive analysis of the energy-dependent phase shifts in the X-e scattering problem has demonstrated~\cite{ImamogluSchmidt} that their low-energy behavior is well captured by classical charge-dipole interactions, in agreement with our results and conclusions. 

The developed microscopic theory for XPs does not rely on any particular model for the screening of the Coulomb interactions. We have chosen the 2D Coulomb potential only due to the significant simplifications it provides in numerical calculations. This should predict the excited states in the excitonic spectrum reasonably well, while the ground state is instead well described by the Keldysh potential~\cite{Keldysh1,Keldysh2,Keldysh3} that properly treats the dielectric screening in layered structures~\cite{TMDCEx1,TrionExperiment4} (including TMDC monolayers embedded in hBN or deposited on a dielectric substrate). However, the universal classical charge-dipole behavior of the calculated X-e interactions suggests a natural generalization of our results to other potentials, including the Keldysh one. For instance, the X-e interactions can be approximated by modified charge-dipole interactions~(\ref{VintAnalytic}), which depend only on the polarizability of the excitonic state $\alpha$, and on the length $a_0$, which together incorporate the details of the screening and band structure. The polarizability can either be obtained from first-principle calculations~\cite{StarkTheory1, StarkTheory2,StarkTheory3,StarkTheory4,ImamogluSchmidt} of the quadratic Stark effect for the ground excitonic state, or it can be extracted from optical absorption measurements in the presence of a direct-current electric field. The value of $a_0$ can then be used as a free parameter to fit the splitting between the XP branches which, at low doping, is equal to the trion binding energy.    

%The considered model is %thoughtfully 
%chosen to avoid the exchange physics that  appears if the photo-excited electron and the electrons of the Fermi sea are indistinguishable. 

%In this work, we have considered the experimentally relevant scenario where the photo-excited electron is distinguishable from those of the Fermi sea.
%Its role 

The scenario where the photo-excited electron is indistinguishable from those of the Fermi sea has been addressed
%
%The opposite scenario has previously been addressed only 
within three-particle physics that is valid in the low doping regime, $\epsilon_\mathrm{F}\ll\epsilon_\mathrm{T}$~\cite{TrionEnergyRecent}. In this case, exchange physics limits correlations, and trions (intravalley ones in TMDCs or spin-singlet ones in semiconductor QWs) are only stable if there is a strong imbalance between the masses of electrons and holes, or in the presence of a strong magnetic field~\cite{MacDonaldTrion,Dzyubenko1,Dzyubenko2,TrionQHTheory1,TrionQHTheory2}. While experiments confirm this prediction, the interplay of exchange and polaronic physics is still poorly understood. This question can be naturally addressed based on our proposed variational ansatz for the XP state, Eq.~\eqref{P}, but this is outside the scope of the present work. 

We have focused on the polaronic splitting for the ground excitonic state $|0,0\rangle$ and have considered its coupling to the excited states perturbatively. The splitting for the excited state $|1,0\rangle$ (the other two states $|1,\pm 1\rangle$ are optically dark) that has been recently reported~\cite{2Strion1,2Strion2,2Strion3} is outside the scope of the current work, but the ability of our approach to address this physics appears promising.

Details of the X-e interactions are especially important if they reside in closely spaced layers. The strong interlayer polaronic coupling opens new avenues to control the motion of excitons via the strong Coulomb drag effect~\cite{PolaronDragTheory}, which cannot be captured by standard perturbative approaches~\cite{CDreview}. Moreover, control over the flow of neutral excitons via electric and magnetic fields has recently been experimentally demonstrated~\cite{PolaronDragExp}.   

To conclude, we have microscopically derived X-e interactions with the help of variational and perturbative approaches. The interactions only weakly depend on doping and can be well approximated by the classical charge-dipole potential. We have calculated the doping dependence of the optical conductivity, and demonstrated that this is well captured by a model that uses instead a phenomenological exciton-electron contact potential. This indicates that including the finite range of interactions is not essential to capturing the physics of exciton-polarons. 

%We compared the doping dependence of the absorption with our previous results obtained for the phenomenological contact potential and found that the doping dependence of the spitting between attractive and repulsive XPs and their broadening can be well approximated by the contact interactions. However, the redistribution of the spectral (oscillator) weights with doping is strongly underestimated, indicating that the finite range of interactions is important.

\acknowledgments 
We acknowledge useful discussions with Francesca Maria Marchetti, Jonathan Keeling and David Reichman. We acknowledge support from the Australian Research Council Centre of Excellence in Future Low-Energy Electronics Technologies (CE170100039). JL is also supported through the Australian Research Council Future Fellowship FT160100244.

\paragraph*{Note added.---}During the preparation of this work we became aware of a recent series of related papers~\cite{Manolatou1,Manolatou2,Manolatou3}. These papers address the trion/exciton-polaron physics with coupled two- and four-particle correlation functions, and they are both consistent with and complementary to our microscopic theory based on the variational approach.

\appendix

\section{Perturbation theory}
\label{AppPerturbationTheory} This appendix works through the perturbative approach to dealing with the contributions of excited states, $\phi_\nu$ and $\chi_{\nu \vec{k}\vec{k}'}$, in the variational equations~(\ref{Variational2}). If we neglect interactions between the excited states themselves (since they are of second order in $\epsilon_\mathrm{T}/\epsilon_\mathrm{X}$), then these amplitudes become
\begin{subequations}
\begin{align}
\phi_\nu& =\frac{\sum_{\vec{k} \vec{k'}}  \Lambda_{\vec{k}-\vec{k'}}^{\nu 0} \chi_{0 \vec{k}\vec{k}'} }{E^\mathrm{XP}-E^\mathrm{X}_\nu}\,, 
\\ 
\chi_{\nu \vec{k}\vec{k}'} & =\frac{ \Lambda_{\vec{k}'-\vec{k}}^{\nu 0} \phi_0  +  \sum_{\bar{\vec{k}}} \Lambda_{\bar{\vec{k}}-\vec{k}}^{\nu 0}  \chi_{0 \bar{\vec{k}} \vec{k}'}}{E^\mathrm{XP}-E^\mathrm{X}_\nu-\epsilon_{\vec{k}\vec{k}'}^\mathrm{FS}}\,. 
\end{align} 
\end{subequations}
Note that, in this section, momenta satisfy the conditions $k,\,\bar{k}>k_\mathrm{F}$ and $k',\,\bar{k}'<k_\mathrm{F}$ due to the presence of the Fermi sea.  After inserting the above amplitudes into the equations for $\phi_{0} \equiv \phi $ and $ \chi_{0\vec{k}\vec{k}'} \equiv \chi_{ \vec{k}\vec{k}'}$, which correspond to the ground state $|\nu\rangle=|0,0\rangle$, we obtain  
\begin{subequations}
\label{VariationalS}
\begin{align}
&\left(\Delta E^\mathrm{XP}_0-\sum_{\vec{k}'} V_{\vec{k}' \vec{k}' \vec{k}'}\right)\phi =  \sum_{\vec{k}\vec{k}'} V_{\vec{k} \vec{k}' \vec{k}'}\chi_{\vec{k}\vec{k}'}\,, \\ 
&\left(\Delta E^\mathrm{XP}_0- \epsilon^\mathrm{FS}_{\vec{k}\vec{k}'}\right)\chi_{\vec{k}\vec{k'}} = S_{\vec{k}\vec{k}'} + V_{\vec{k}' \vec{k} \vec{k}'} \phi+\sum_{\bar{\vec{k}}} V_{\bar{\vec{k}} \vec{k} \vec{k}'}\chi_{\bar{\vec{k}}\vec{k'}}\,.
\end{align}
\end{subequations}
Here, $\Delta E^\mathrm{XP}_0=E^\mathrm{XP}-E^\mathrm{X}_0$ is the energy of the exciton-polaron state $E^\mathrm{XP}$ defined with respect to the energy of the excitonic ground state $E^\mathrm{X}_0$. These expressions~(\ref{VariationalS}) coincide with the system of equations~(\ref{Variational3}) presented in the main text, except for the additional term
\begin{equation}
\label{Sfactor}
S_{\vec{k}\vec{k}'}=\sum_{\nu\,\neq\,0} \frac{\Lambda^{0\nu}_{\vec{k}'-\vec{k}}}{E^\mathrm{XP}-E^\mathrm{X}_\nu} \sum_{\bar{\vec{k}}\bar{\vec{k}}'}\Lambda^{\nu0}_{\bar{\vec{k}}-\bar{\vec{k}}'}\chi_{\bar{\vec{k}}\bar{\vec{k}}'}\,. 
\end{equation}
However, we now prove that this term is zero. %and is zero $S_{\vec{k}\vec{k}'}=0$. 
First we note that, due to rotational symmetry, $\chi_{\bar{\vec{k}}\bar{\vec{k}}'}$ above depends only on the relative angle $\theta_{\bar{\vec{k}} \bar{\vec{k}}'}$ between momenta $\bar{\vec{k}}$ and $\bar{\vec{k}}'$. Furthermore, the matrix element $\Lambda^{\nu0}_{\bar{\vec{k}}-\bar{\vec{k}}'}$ can be factorized as $\Lambda^{\nu0}_{\bar{\vec{k}}-\bar{\vec{k}}'}=\Pi_l\,e^{-i l\theta_{\bar{\vec{k}}'}} F(\bar{k},\bar{k}',\theta_{\bar{\vec{k}} \bar{\vec{k}}'})$ [see Eq.~(\ref{NewLambdaEq}) later].  Here, $\Pi_{l}=\sin(\pi |l|/2)$ is the parity factor that is nonzero for only odd values of $l$, and $F(\bar{k},\bar{k}',\theta_{\bar{\vec{k}} \bar{\vec{k}}'})$ depends \emph{only} on the relative angle between momenta $\theta_{\bar{\vec{k}} \bar{\vec{k}}'}$. If we shift the angular integration in Eq.~(\ref{Sfactor}) to $d\theta_{\bar{\vec{k}}}d\theta_{\bar{\vec{k}}'}\rightarrow d\theta_{\bar{\vec{k}}\bar{\vec{k}}'} d\theta_{\bar{\vec{k}}'}$ and integrate over $d\theta_{\bar{\vec{k}}'}$, then we find that only excited states with $l=0$ can contribute. However, their contribution in fact vanishes due to the parity factor $\Pi_0=0$. As a result, the term $S_{\vec{k}\vec{k}'}$ is zero and Eq.~(\ref{VariationalS}) reduces to Eq.~(\ref{Variational3}) in the main body. 

\section{The spectrum of excitonic states}
\label{AppExcitonicSpectrum}
In this appendix, we briefly review the analytic solution to the 2D Coulomb problem~\cite{Coulomb2D1,Coulomb2D2,Coulomb2DPortnoi}. The spectrum consists of a discrete component that corresponds to bound excitons, and also a continuum of unbound electron-hole pairs. 
%
%The Appendix presents the spectrum of 2D Coulomb problem that admits the analytical solution~\cite{Coulomb2D1,Coulomb2D2,Coulomb2DPortnoi}. Its discrete part of the spectrum corresponds to bound excitons while the continuous one to unbound electron-hole pairs. 
%
The discrete excitonic spectrum is labeled by radial $n=0,\,1,\,2,\,...$ and orbital $l=-n,\,...,\,n$ quantum numbers. These can be combined into a single index $|\nu\rangle=|n,l\rangle$ which is used in the main text of the paper. The corresponding binding energies, $E^\mathrm{X}_\nu=-\varkappa_{\nu}^2 \epsilon_\mathrm{X}$, do not depend on $l$ and therefore feature the \emph{accidental degeneracy} known for Coulomb problems~\cite{Coulomb2DPortnoi}. Here, $\varkappa_{\nu}=1/(2n+1)$, and $\epsilon_\mathrm{X}=m e^4/\hbar^2\kappa^2$ is the binding energy for the ground excitonic state.  The associated real-space wave functions, $C^{nl}_\vec{r}=R_{nl}(r) \Phi_l(\theta_\vec{r})$, can be written as a product of the angular part, $\Phi_l(\theta_\vec{r})=e^{i l\theta_\vec{r}}/\sqrt{2\pi}$, and the radial part, $R_{nl}(r)$, given by 
\begin{equation}
R_{nl}(r)=\frac{2\varkappa_{\nu} N_{nl}}{a_\mathrm{X}}\rho^{|l|} L_{n-|l|}^{2|l|}(\rho)e^{-\rho/2}\,.
\end{equation}
Above, $\rho=2\varkappa_{\nu} r/a_\mathrm{X}$, and $a_\mathrm{X}=\hbar^2 \kappa/m e^2$ is the average separation between electron and hole in the excitonic ground state. In addition, $L_{n}^a(x)$ are generalized Laguerre polynomials and $N_{nl}$ is the normalization coefficient where 
\begin{equation}
N_{nl}^2=\varkappa_\nu\frac{(n-|l|)!}{(n+|l|)!}\,.
\end{equation}

The continuous spectrum for unbound electron-hole pairs is labeled by the dimensionless absolute momentum $q$ and the orbital quantum number $l=-\infty,\,...,\,\infty$. Again, these can be combined into a single index $|\nu\rangle=|q,l\rangle$, and the corresponding energies are $E^\mathrm{X}_\nu=q^2 \epsilon_\mathrm{X}$. The wave functions, $C^{q l}_\vec{r}=R_{q l}(r) \Phi_l(\theta_\vec{r})$, can again be presented as a product where now the radial part, $R_{ql}(r)$, is given by  
\begin{equation}
R_{q l}(r)=\frac{2 q  N_{ql}}{a_\mathrm{X}}\rho^{|l|} {}_1 F_1\hspace{-0.2em}\left(\frac{i}{2q}+ |l|+\frac{1}{2},2|l|+1, i \rho\right)\hspace{-0.1em}e^{-i\rho/2}\,.
\end{equation}
Here, $\rho=2q r/a_\mathrm{X}$, and ${}_1 F_1 (a,b,z)$ is the Kummer confluent hypergeometric function. The normalization coefficient, $N_{ql}$, is written as
\begin{equation}
%N_{ql}= \sqrt{\frac{\pi}{q}}\frac{1}{\sqrt{1+e^{-\pi/q}}} \frac{S_{ql}}{(2|l|)!},\nn\\
N_{ql}= \frac{S_{ql}}{(2|l|)!}\sqrt{\frac{\pi/q}{1+e^{-\pi/q}}}\,,
\end{equation}
where $S_{ql}$ is given by $S_{q0}=1$ for $l=0$ and %
\begin{equation}
S_{ql}=\prod_{s\,=\,1}^{|l|} \sqrt{\left(s-\frac{1}{2}\right)^2+\frac{1}{4 q^2}}
\end{equation}
otherwise. The wave functions for bound excitonic states and unbound electron-hole pairs determine the X-e interactions via the matrix element for X-e scattering.  This matrix element, $\Lambda_{\vec{k}-\bar{\vec{k}}}^{\nu \bar{\nu}}$, is presented below in App.~\ref{AppMatrixElement}. 
 
\begin{figure}[t]
	\includegraphics[width=.9\columnwidth]{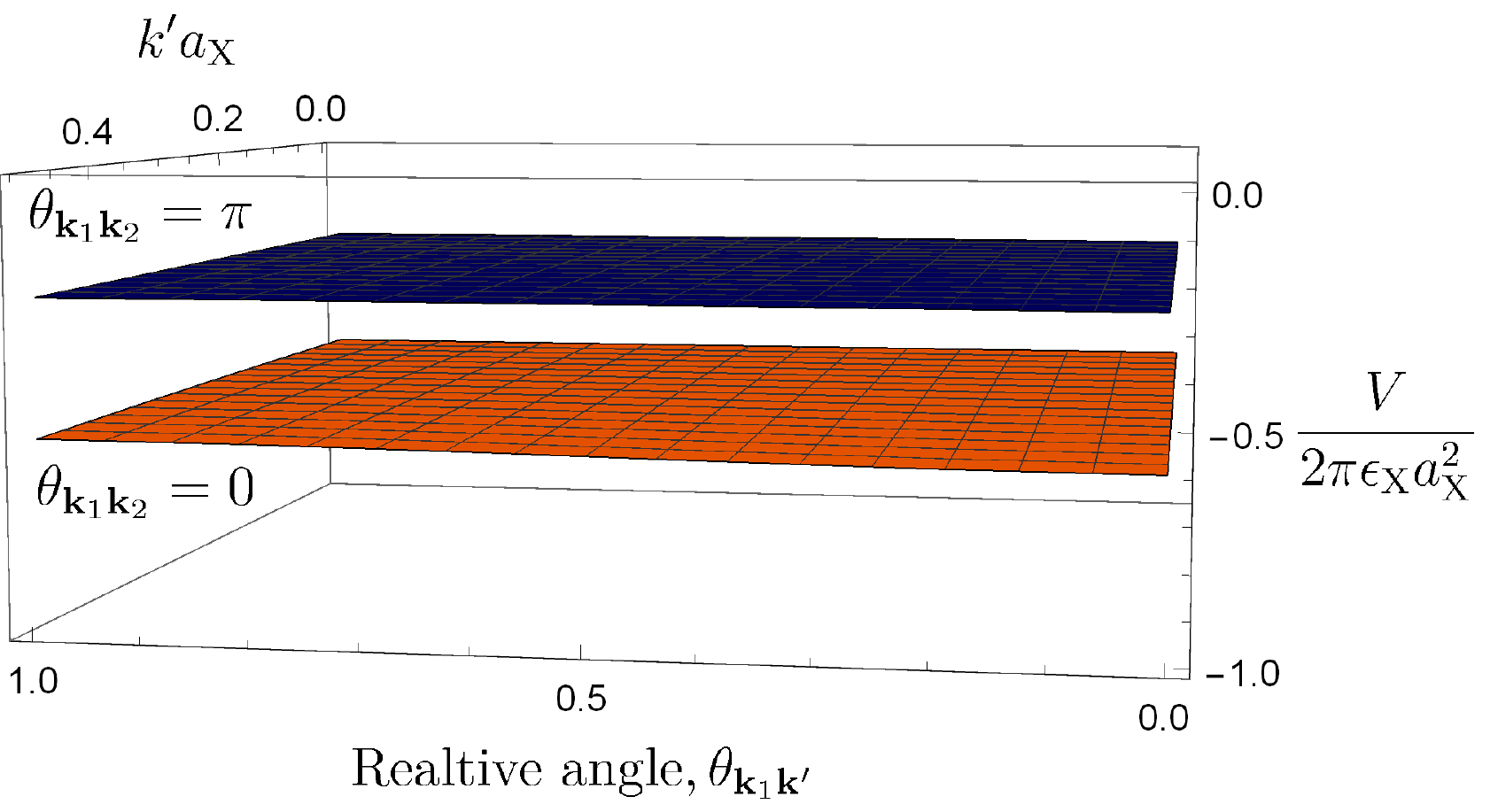}
	\caption{Dependence of the nonlocal interaction, $V_{\vec{k}_1 \vec{k}_2 \vec{k}'}$, on both the magnitude $k'a_\mathrm{X}$ of the Fermi-hole momentum and its relative angle $\theta_{\vec{k}_1\vec{k}'}$. It is calculated for the momenta magnitudes, $k_1 a_\mathrm{X}=k_2 a_\mathrm{X}=1$, and for two values of the relative angle between them, $\theta_{\vec{k}_1\vec{k}_2}=0$ and $\pi$. \label{FigFermiHole}} 
\end{figure}

\begin{figure}[t]
	\includegraphics[width=.8\columnwidth]{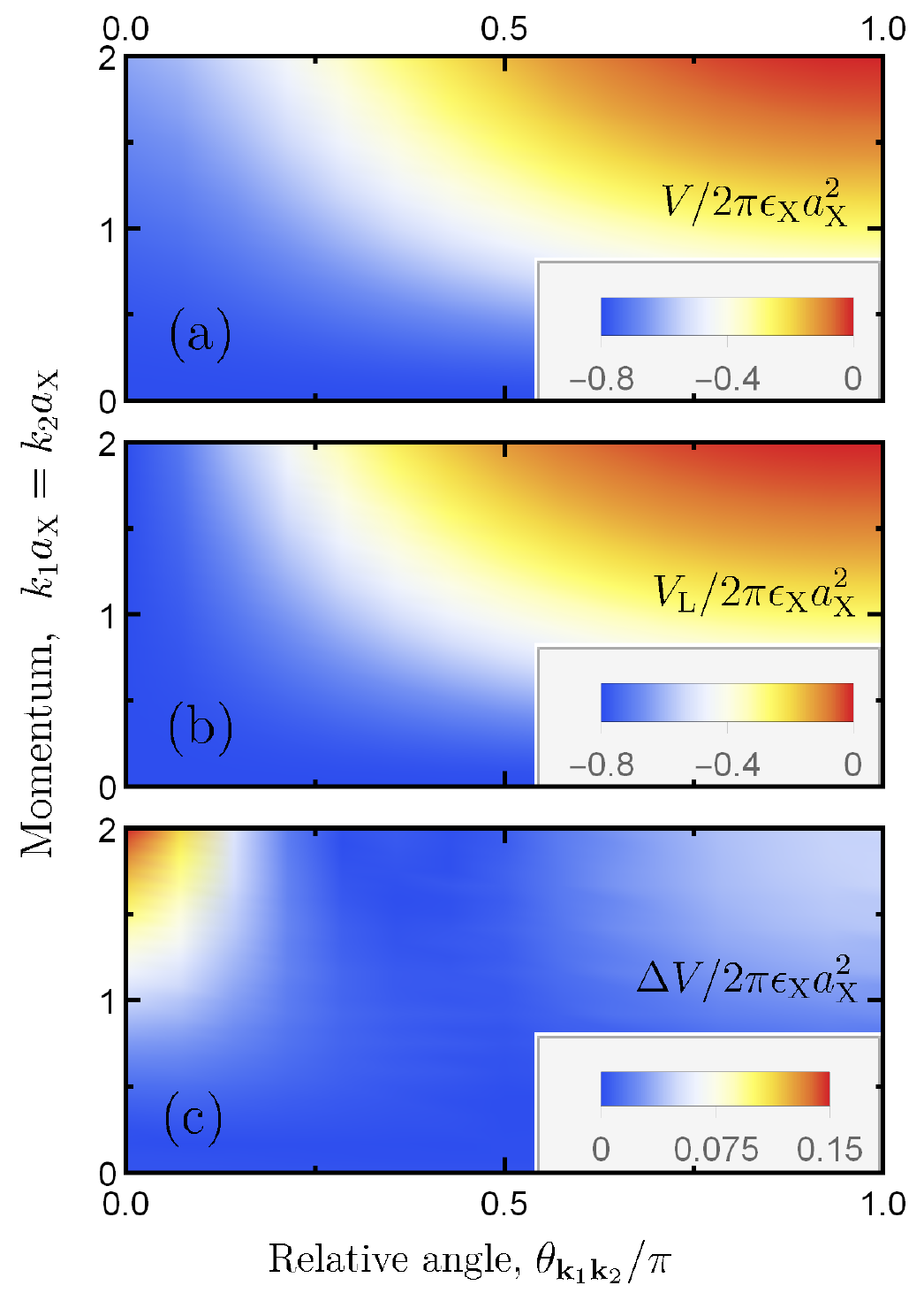}
	\caption{Nonlocal interactions $V_{\vec{k}_1\vec{k}_2 \vec{0}}$ (a), the local one $V_\mathrm{L}(\vec{q})$ with $\vec{q}=\vec{k}_1-\vec{k}_2$ (b), and the discrepancy between them $\Delta V_{\vec{k}_1\vec{k}_2}=|V_{\vec{k}_1\vec{k}_2\vec{0}}-V_\mathrm{L}(\vec{q})|$ (c). Their dependencies on the momenta magnitudes $k_1a_\mathrm{X}= k_2a_\mathrm{X}$ and the relative angle $\theta_{\vec{k}_1\vec{k}_2}/\pi$ are depicted for $\epsilon_\mathrm{F}=0$. \label{FigNonLocalNoDoping}} 
\end{figure}

\section{Scattering matrix element}
\label{AppMatrixElement}
This appendix discusses the matrix element, $\Lambda_{\vec{k}-\bar{\vec{k}}}^{\nu \bar{\nu}}$, of X-e scattering with transferred momentum, $\vec{k}-\bar{\vec{k}}$. The exact expression for the scattering matrix element is 
\begin{equation}
\Lambda^{\nu\bar{\nu}}_{\vec{k}-\bar{\vec{k}}}= U_{\vec{k}-\bar{\vec{k}}}\int d\vec{ r}\,(C_\vec{r}^\nu)^*C_\vec{r}^{\bar{\nu}}\,2 i \sin\left[\frac{(\vec{k}-\bar{\vec{k}})\cdot\vec{r}}{2}\right].
\end{equation}
By employing the explicit form of the spectrum from the 2D Coulomb problem, presented in App.~\ref{AppExcitonicSpectrum}, this matrix element can be rewritten as follows: 
\begin{equation}
\label{NewLambdaEq}
\Lambda^{\nu\bar{\nu}}_{\vec{k}-\bar{\vec{k}}}=i \frac{2\pi e^2}{\kappa} \Pi_{l-\bar{l}} e^{ i (\bar{l}-l) \theta_{\bar{\vec{k}}}} A^{\bar{l}-l}(k,\bar{k},\theta_{\vec{k} \bar{\vec{k}}}) M^{\nu\bar{\nu}}_{|\vec{k}-\bar{\vec{k}}|}\,,
\end{equation}
where $\theta_{\vec{k} \bar{\vec{k}}}$ is the relative angle between $\vec{k}$ and $\bar{\vec{k}}$. Here, the parity factor $\Pi_l$, the angle factor $A(k,\bar{k},\theta_{\vec{k}\bar{\vec{k}}})$, and the integral $M_q^{\nu \bar{\nu}}$, are given by%\vspace{-0.1em}
\begin{equation}
\begin{split}
\Pi_l&=\sin\left(\frac{ \pi|l|}{2}\right), \quad\quad A(k,\bar{k},\theta_{\vec{k}\bar{\vec{k}}})=\frac{k e^{i \theta_{\vec{k}\bar{\vec{k}}}}-\bar{k}}{|\vec{k}-\bar{\vec{k}}|}\,, \\ 
M_{q}^{\nu\bar{\nu}}&= \int_0^\infty rdr R_{\nu}(r)R_{\bar{\nu}}(r) \frac{2 }{q}J_{|l-\bar{l}|}\left(\frac{q r}{2}\right).
\end{split}
\end{equation}
Above, $R_\nu(r)$ is the radial part of the wave function,~$C_\vec{r}^\nu$, and $J_n(z)$ is the $n^{th}$-order Bessel function of the first kind. Importantly, $\Pi_{l-\bar{l}}$ is nonzero only if $l-\bar{l}$ is odd, which implies that the states $|\nu\rangle$ and $|\bar{\nu}\rangle$ have opposite parities. For this reason, the matrix element of elastic X-e scattering is zero, $\Lambda_{\vec{k}-\bar{\vec{k}}}^{\nu \nu}=0$, and hence X-e interactions appear only at second order in perturbation theory. %, but not at first order. 
(We discuss that further in App.~\ref{AppXescattering}.) This argument is based solely on the symmetries of the excitonic states and is insensitive to details of the electron-hole interaction, $U_\vec{q}$. The latter impacts only on the magnitude of $M^{\nu \bar{\nu}}_q$.

\begin{figure}[t]
	\includegraphics[width=.8\columnwidth]{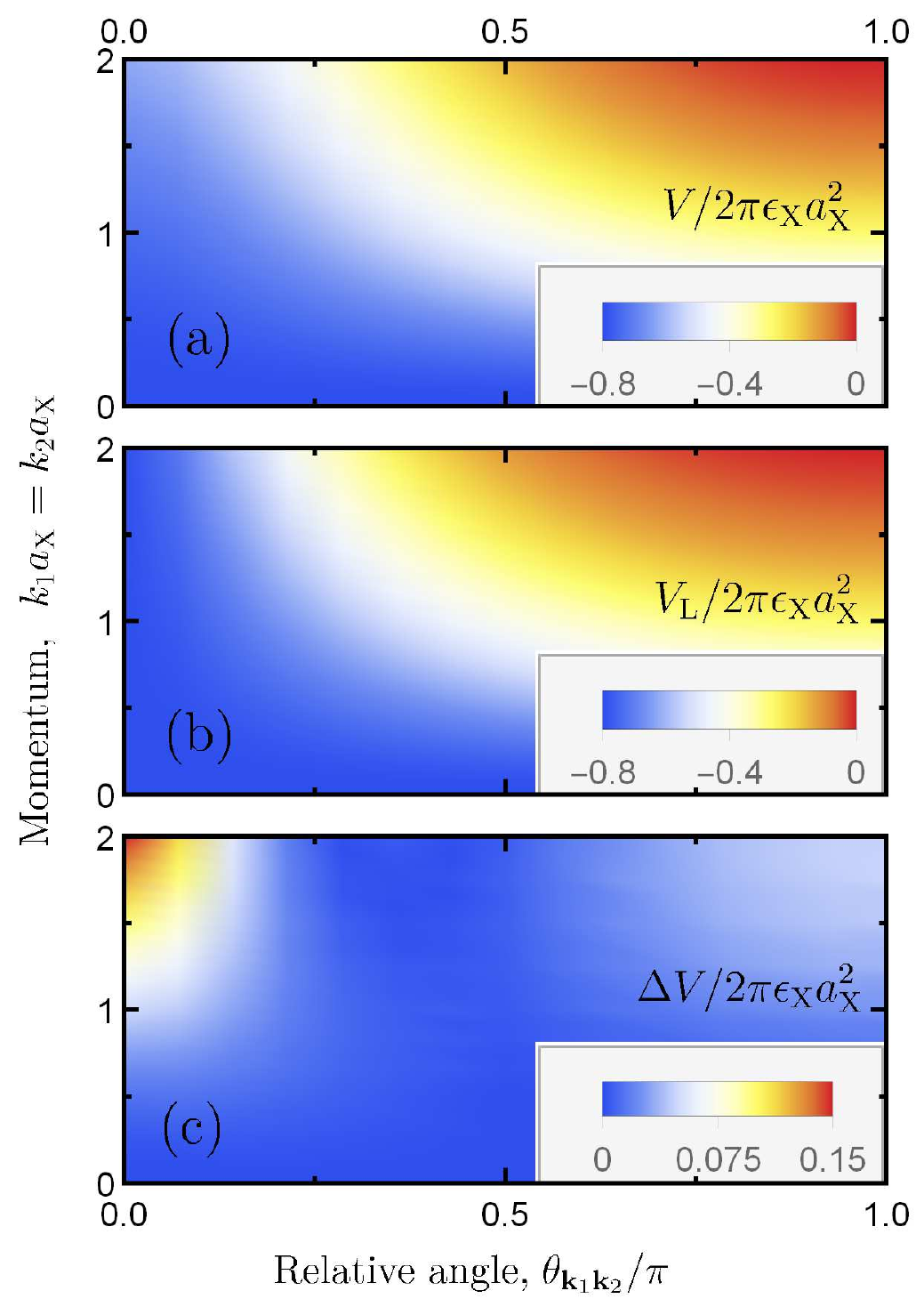}
	\caption{Nonlocal interactions $V_{\vec{k}_1\vec{k}_2 \vec{0}}$ (a), the local one $V_\mathrm{L}(\vec{q})$ with $\vec{q}=\vec{k}_1-\vec{k}_2$ (b), and the discrepancy between them $\Delta V_{\vec{k}_1\vec{k}_2}=|V_{\vec{k}_1\vec{k}_2\vec{0}}-V_\mathrm{L}(\vec{q})|$ (c). Their dependencies on the momenta magnitudes $k_1a_\mathrm{X} = k_2a_\mathrm{X} $ and the relative angle $\theta_{\vec{k}_1\vec{k}_2}/\pi$ are depicted for $\epsilon_\mathrm{F}=0.12\,\epsilon_\mathrm{X}$. \label{FigNonLocalDoping}}
\end{figure}

\section{Nonlocal exciton-electron interactions}
\label{AppNonlocal}
This appendix presents a comparison between the microscopically derived nonlocal exciton-electron interaction, $V_{\vec{k}_1 \vec{k}_2 \vec{k}'}$ (\ref{Vint}), and its local approximation, $V_\mathrm{L}(\vec{q})$ (\ref{Vintlocal}) with $\vec{q}=\vec{k}_1-\vec{k}_2$. The nonlocal interaction potential $V_{\vec{k}_1 \vec{k}_2 \vec{k}'}$ is doping dependent, and a function of three momenta $k_1,\,k_2,\,k'$ and two relative angles $\theta_{\vec{k}_1 \vec{k}_2},\,\theta_{\vec{k}_1 \vec{k}'}$. 

First, we notice that the dependence of the interaction $V_{\vec{k}_1 \vec{k}_2 \vec{k}'}$ on the momentum of the redundant Fermi hole, $k'$ and $\theta_{\vec{k}_1 \vec{k}'}$, is extraordinarily weak for any $k_1$ and $k_2$ and $\theta_{\vec{k}_1 \vec{k}_2}$.  (Recall that this hole is not directly involved in the scattering event.)  We illustrate this in Fig.~\ref{FigFermiHole} for $k_1 a_\mathrm{X}=k_2a_\mathrm{X}=1$ and for two values of their relative angle, $\theta_{\vec{k}_1\vec{k}_2}=0$ and $\pi$. As a result, the nonlocal potential is very well approximated by its value at $\vec{k}'=\vec{0}$. 

A preliminary comparison between the interactions, $V_{\vec{k}_1 \vec{k}_2 \vec{0}}$ and $V_\mathrm{L}(\vec{q})$, shows that the discrepancy between them,  $\Delta V_{\vec{k}_1\vec{k}_2}=|V_{\vec{k}_1\vec{k}_2\vec{0}}-V_\mathrm{L}(\vec{q})|$, achieves a maximum at $k_1a_\mathrm{X}=k_2a_\mathrm{X}$. Their dependencies on these magnitudes of momenta and the angle between them $\theta_{\vec{k}_1\vec{k}_2}$ are presented in Figs.~\ref{FigNonLocalNoDoping}~and~\ref{FigNonLocalDoping} for $\epsilon_\mathrm{F}/\epsilon_\mathrm{X}=0$ and $0.12$ (respectively). Due to the ordering of energy scales, $\epsilon_\mathrm{T}\ll\epsilon_\mathrm{X}$, the length scale of the polaronic correlations is estimated to be $a_\mathrm{T}\approx a_\mathrm{X} \sqrt{\epsilon_\mathrm{X}/\epsilon_\mathrm{T}}\approx 3 a_\mathrm{X}$, which corresponds to $q a_\mathrm{X}\sim 1/3$. It is clear that the local potential quite well approximates X-e interactions in the regime where $q a_\mathrm{X}\lesssim 1$. Therefore, the nonlocal nature of these interactions is important only at $q a_\mathrm{X}\gtrsim1$, corresponding to an X-e separation that is smaller than the exciton size. % $r a_\mathrm{X}$. 

\section{Exciton-electron scattering}
\label{AppXescattering}
In this appendix, we connect the matrix element $\Lambda_{\vec{q}}^{\nu \bar{\nu}}$ presented in App.~\ref{AppMatrixElement} with the amplitude for X-e scattering. We additionally argue that the potential derived in Ref.~\cite{XPnew1} appears to be incorrect. If we treat the exciton as a structureless bosonic particle, then its scattering is solely determined via interactions with electrons, $V_\mathrm{S}(\vec{q})$. This can be expressed by the scattering amplitude within the Born approximation: 
\begin{equation}
\label{VSA1}
V_\mathrm{S}(\vec{q})=\langle f_0| H_\mathrm{X\text{-}e}|i_0 \rangle\,.
\end{equation}
Here, $|i_\nu\rangle=X^\dagger_{\nu \vec{p}_\mathrm{X}} f^\dagger_{\vec{p}_\mathrm{e}}|\mathrm{g}\rangle$ and  $|f_\nu\rangle=X^\dagger_{\nu \vec{p}_\mathrm{X}+\vec{q}} f^\dagger_{\vec{p}_\mathrm{e}-\vec{q}}|\mathrm{g}\rangle$ are the excitonic states before and after scattering, with label $\nu$, and $|\mathrm{g}\rangle$ is the unexcited Fermi sea.  The electron $f^\dagger_{\vec{p}_\mathrm{e}}$ and exciton $X^\dagger_{\nu \vec{p}_\mathrm{X}}$ operators have been respectively introduced in Eqs.~(\ref{H0})~and~(\ref{X}).  Importantly, here, the exciton is assumed to reside in the ground state, $\nu=0$, which indicates that the scattering is elastic.

The scattering matrix element~(\ref{VSA1}) can be evaluated microscopically by recalling the composite nature of the exciton and substituting $H_{\mathrm{X\text{-}e}}$ with $H_{\mathrm{e\text{-}f}}+H_{\mathrm{h\text{-}f}}$ [the first and second terms in Eq.~\eqref{eq:HC}]. A straightforward calculation $V_\mathrm{S}(\vec{q})=\Lambda_{\vec{q}}^{0 0}$ then yields
\begin{equation}
\begin{split}
\label{VSA2}
V_\mathrm{S}(\vec{q})&=\langle f_0|\big(H_{\mathrm{e\text{-}f}}+H_{\mathrm{h\text{-}f}}\big)|i_0 \rangle \\
&=U_\vec{q}\sum_\vec{p} \big(C_\vec{p}^0\big)^* \left(C_{\vec{p}-\vec{q}/2}^0- C_{\vec{p}+\vec{q}/2}^0\right) = 0\,.
\end{split}
\end{equation}
Above, $\Lambda_{\vec{q}}^{\nu \bar{\nu}}$ is the scattering matrix element between states $\nu$ and $\bar{\nu}$ which is discussed in App.~\ref{AppMatrixElement}. We see that $V_\mathrm{S}(\vec{q})=0$ because  $\Lambda_{\vec{q}}^{\nu \bar{\nu}}$ is nonzero only between states of different parities (explained in App.~\ref{AppMatrixElement}).  Thus, we conclude that there is no contribution to X-e interactions from first-order perturbation theory. The vanishing scattering amplitude, $V_\mathrm{S}(\vec{q})$, reflects the fact that X-e interactions appear due to the polarization of the exciton by the electric field of the electron. This process appears at second order in perturbation theory, with no contribution at first order.

\bibliography{main.bib}

\end{document}